\documentclass{aa}

\usepackage{graphicx}
\usepackage{txfonts}
\usepackage{lipsum}
\usepackage{subcaption}         
\usepackage{lscape}             
\usepackage{placeins}           

\usepackage{amsmath}
\allowdisplaybreaks[4]
\usepackage{amssymb}

\usepackage{float}
\usepackage{leftidx}
\usepackage{CJK}
\usepackage{esint}
\usepackage{bm}
\usepackage{color}
\usepackage{tabu}
\usepackage{booktabs}
\usepackage{threeparttable}
\usepackage{multirow}
\usepackage{longtable}
\usepackage{rotating}
\usepackage{makecell}
\usepackage{soul}

\usepackage{url}
\usepackage[colorlinks, linkcolor=blue, anchorcolor=blue, citecolor=blue, filecolor=blue, menucolor=blue, runcolor=blue, urlcolor=blue]{hyperref}

\begin{document}
  
  \title{Simulations for the evolution of the chemical clock HC$_3$N/N$_2$H$^+$ in high-mass star-forming regions}
  \author{Yao Wang\inst{1}\and Fujun Du\inst{1,2}\and Youxin Wang\inst{3}\and Hongchi Wang\inst{1,2} \and Jiangshui Zhang\inst{4}}
  \institute{Purple Mountain Observatory, Chinese Academy of Sciences, 10 Yuanhua Road, 210023 Nanjing, PR China; \email{wangyao@pmo.ac.cn}
  \and
University of Science and Technology of China, 96 Jinzhai Road, 230026 Hefei, PR China; 
        \and
Max-Planck-Institut f{\"u}r Radioastronomie, Auf dem H{\"u}gel 69, 53121 Bonn, Germany; 
        \and
Center for Astrophysics, Guangzhou University, 230 Wai Huan Xi Road, 510006 Guangzhou, PR China
  }
  \date{Received date month year / Accepted date month year}
  
  
    \abstract{From observations, column density ratios or integrated intensity ratios between some species exhibit monotonic increase or decrease along with the evolution of high-mass star-forming regions (HMSFRs). Such ratios are defined as chemical clocks, which can be used to constrain the evolutionary stage. }
    {We performed chemical simulations to reproduce the observed column density ratio of HC$_3$N/N$_2$H$^+$ and the abundances of these two species across various evolutionary stages in HMSFRs. Simultaneously, we identified the chemical processes responsible for the observed time-dependent trends in these stages. }
    {Our simulations utilized the astrochemical code Nautilus and the existing 1D models of HMSFRs that cover four evolutionary stages, accompanied by variations in density and temperature throughout the entire evolution. In addition, to maintain a steady increase in density and temperature over time as predictions of the global hierarchical collapse scenario, we adjusted parameters such as density, temperature, and time spent in each evolutionary stage. }
    {When averaging over large spatial scales, the best model produced successfully matches the observed column density ratio of HC$_3$N/N$_2$H$^+$ and the abundances of the species involved at specific times for each evolutionary stage; that is, the late high-mass starless core stage, the early high-mass protostellar object stage, and the early ultracompact H{\sc ii} stage. HC$_3$N is mainly affected by the warm carbon-chain chemistry (WCCC) and its own thermal desorption, while N$_2$H$^+$ is primarily influenced by the thermal desorption of N$_2$, CO, CH$_4$, NH$_3$, and H$_2$O followed by dissociative recombination and ion-molecule reactions. }
    {The results obtained from the best-fitting model timescales broadly agree with statistical estimates. However, a continuous increasing ratio of HC$_3$N/N$_2$H$^+$ throughout the entire evolution of HMSFRs is not acquired. Some observed ratios between adjacent stages overlap, which could be induced by observational uncertainties (such as those in deriving column densities and abundances, clump classification, and systematic effects), or indicate that the evolution of HC$_3$N/N$_2$H$^+$ may not strictly monotonically increase throughout the entire evolution. Based on our best-fit model, we further examined other 350 ratios involving 27 species, and 178 ratios exhibit an increasing or decreasing evolutionary trend around the best-fit timescales of HC$_3$N/N$_2$H$^+$. Among them, 157 ratios are observable and could be considered as candidate chemical clocks. Our results indicate that 1D models with abrupt jumps in physical parameters have reached their limits in terms of the insights they can provide, and more sophisticated models need to be adopted. }

   
   \maketitle

  \section{Introduction} \label{section1}
  
  \par{}High-mass stars are important for the energy generation and the life cycle of the interstellar medium (ISM) in the Universe through radiation, outflows, stellar wind, and supernova events \citep{2018ARA&A..56...41M}. Thus, it is crucial to track the evolution of high-mass stars throughout their whole lifetime. However, currently, high-mass star formation is still unclear compared with low-mass star formation, because of the difficulties in observation induced by the small number of targets \citep[$\sim 3\%$;][]{2001MNRAS.322..231K, 2023ApJS..264...48W}, large distances from Earth \citep[several kiloparsecs;][]{2018MNRAS.473.1059U, 2021A&A...648A..66G}, and typically short evolutionary timescales \citep[the order of 10$^5$ yr;][]{2011ApJ...730L..33M, 2018A&A...616A.101K, 2018ARA&A..56...41M, 2021A&A...648A..66G}. Nevertheless, based on the observed properties, the evolution of high-mass star-forming regions (HMSFRs) can be roughly divided into four distinct stages, including the high-mass starless core (HMSC) stage, the high-mass protostellar object (HMPO) stage, the hot molecular core (HMC) stage, and the ultracompact H{\sc ii} (UCH{\sc ii}) stage, respectively \citep[e.g.,][]{2007prpl.conf..165B, 2007ARA&A..45..481Z, 2014A&A...563A..97G, 2015A&A...579A..80G, 2015ApJS..219....2J, 2021A&A...648A..66G}. The HMSCs are molecular cores of cold dense gas and dust with temperatures of $10-20\,\mathrm{K}$ \citep{2006A&A...450..569P, 2007ARA&A..45..481Z, 2021A&A...648A..66G}, which are close to isothermal and radiate mostly at millimeter/submillimeter wavelengths; they commonly belong to infrared dark clouds \citep[IRDCs;][]{2005ApJ...634L..57S, 2006ApJ...641..389R}. Subsequently, HMPOs are formed: gravitational collapse generates protostars that are still accreting mass \citep[$>8\,M_{\sun}$;][]{2007prpl.conf..165B, 2014A&A...563A..97G, 2015A&A...579A..80G}, and that can radiate at mid-infrared wavelengths and heat the surrounding environment to become non-isothermal \citep{2002ApJ...566..945B, 2016A&ARv..24....6B}. Next, a higher temperature \citep[$>100\,\mathrm{K}$;][]{2006ApJ...641..389R, 2018A&A...617A..60S, 2021A&A...648A..66G} prompts the formation of HMCs, accompanied by the increased gaseous chemical complexity of various molecules but similar physical conditions compared to HMPOs \citep{2014A&A...563A..97G, 2015A&A...579A..80G}. Finally, UCH{\sc ii} regions are created by strong radiation from protostars, and are associated with ionized gas, destroyed complex molecules, and free-free emission at centimeter wavelengths \citep{2010ApJ...719..831P, 2013ApJ...766..114S, 2013MNRAS.435..400U, 2018A&A...617A..60S, 2021A&A...648A..66G}. Due to the smooth evolution, these stages cannot be clearly distinguished, particularly among HMPO, HMC, and UCH{\sc ii} stages \citep{2014A&A...563A..97G, 2015A&A...579A..80G, 2021A&A...648A..66G}. 
  
  \par{}Chemical evolution is a useful tool for further analyzing high-mass star formation. Diverse chemical compositions across various sources can be used to infer physical properties and subsequently constrain different evolutionary conditions \citep[e.g.,][]{2009ARA&A..47..427H, 2014A&A...563A..97G, 2020ARA&A..58..727J, 2022ApJS..259...30M}. The column density ratios or integrated intensity ratios between different species are usually used to distinguish different evolutionary stages and constrain lifetimes of HMSFRs, which are defined as chemical clocks \citep{2004A&A...422..159W, 2013ApJ...777..157H, 2015A&A...579A..80G, 2015MNRAS.451.2507Y, 2019MNRAS.484.4444U, 2023ApJS..264...48W}. Such ratios commonly increase or decrease monotonically along with the entire evolution of HMSFRs. For instance, \citet{2013ApJ...777..157H} derived that the relative abundance ratio of N$_2$H$^+$/HCO$^+$ increased slightly throughout the evolution according to observed 333 sources belonging to the Millimeter Astronomy Legacy Team 90 GHz (MALT90) survey. \citet{2014PASJ...66..119O} and \citet{2014PASJ...66...16T} proposed that the column density ratios of $N$(c-C$_3$H$_2$)/$N$(C$_2$S), $N$(NH$_3$)/$N$(C$_2$S), $N$(NH$_3$)/$N$(HC$_3$N), and $N$(N$_2$H$^+$)/$N$(C$_2$S) in starless regions could be lower than those in star-forming regions based on the observations of six molecular clouds in Orion A. \citet{2015ApJS..219....2J} estimated that the abundance ratio of HCN/HNC increased from 0.97 in quiescent IRDC cores, then to 2.65 in active IRDC cores, next to 4.17 in HMPOs, and finally to 8.96 in UCH{\sc ii} regions among a total of 78 sources. \citet{2015MNRAS.451.2507Y} suggest that the abundance ratios of N$_2$H$^+$/H$^{13}$CO$^+$ and C$_2$H/H$^{13}$CO$^+$ decrease from massive young stellar objects (MYSOs) to UCH{\sc ii} regions toward 31 clumps from the MALT90 survey. For the statistics of larger samples, based on 3246 high-mass clumps from the MALT90 survey, \citet{2016PASA...33...30R} found that the integrated intensity ratios of HCO$^+$/HNC, HCN/HNC, HCO$^+$/H$^{13}$CO$^+$, HNC/HN$^{13}$C, N$_2$H$^+$/H$^{13}$CO$^+$, and H$^{13}$CO$^+$/SiO increase along with the evolution, while the ratios of HCO$^+$/N$_2$H$^+$, HNC/N$_2$H$^+$, and C$_2$H/N$_2$H$^+$ decrease from quiescent clumps to protostars then increase into UCH{\sc ii} regions. Similarly, \citet{2019MNRAS.484.4444U} propose 13 integrated intensity ratios as candidate chemical clocks among 570 sources from the APEX Telescope Large Area Survey of the Galaxy (ATLASGAL), and that H$^{13}$CN/N$_2$H$^+$, $^{13}$CS/N$_2$H$^+$, HCN/HNC, HCN/N$_2$H$^+$, HC$_3$N/HN$^{13}$C, H$^{13}$CN/HN$^{13}$C, HC$_3$N/N$_2$H$^+$, C$_2$H/N$_2$H$^+$, HCO$^+$/N$_2$H$^+$, and C$_2$H/c-C$_3$H$_2$ increase monotonically, while H$^{13}$CO$^+$/N$_2$H$^+$, H$^{13}$CO$^+$/HN$^{13}$C, and C$_2$H/HN$^{13}$C also exhibit a tendency to increase, except in quiescent clumps. Considering the statistical significance of observed samples and the deviation degree of ratios across different stages, the most promising chemical clocks could include HCO$^+$/HNC, HCN/HNC, $^{13}$CS/N$_2$H$^+$, HCN/N$_2$H$^+$, HC$_3$N/HN$^{13}$C, and HC$_3$N/N$_2$H$^+$. Among these chemical clocks, N$_2$H$^+$, HCO$^+$, HCN, HNC, and C$_2$H are primarily involved. The two ions are mainly affected by the intensities of ionization and radiation field (e.g., cosmic-ray ionization and dissociation, photoionization, and photodissociation), while the other three molecules are mainly influenced by the temperature and the density (e.g., neutral-neutral reaction, ion-molecule reaction, accretion, and desorption). Therefore, these species can be utilized to research the evolution of HMSFRs. 
  
  \par{}Recently, \citet{2023ApJS..264...48W} further proposed a candidate chemical clock, $N$(HC$_3$N)/$N$(N$_2$H$^+$) (hereafter HC$_3$N/N$_2$H$^+$) in HMSFRs, following the integrated intensity ratio of HC$_3$N/N$_2$H$^+$ proposed by \citet{2019MNRAS.484.4444U}. According to their observations concerning 61 UCH{\sc ii} regions via the Institut de Radioastronomie Millimétrique (IRAM) 30 m and the Arizona Radio Observatory (ARO) 12 m telescopes, combined with previous observations toward 17 HMSCs and 28 HMPOs through the Nobeyama 45 m telescope performed by \citet{2019ApJ...872..154T}, the column density ratio increased by a factor of five from $\sim0.136$ at the HMSC stage to $\sim0.303$ at the HMPO stage, and finally to $\sim0.777$ at the UCH{\sc ii} stage (see Table \ref{table3}). The column density ratio at the HMSC stage is comparable to the integrated intensity ratio ($\sim0.1$). However, at the HMPO and UCH{\sc ii} stage, the column density ratio is higher than the integrated intensity ratio by a factor of two or three ($\sim0.13$ and $\sim0.26$, respectively; see Fig. 20 in \citealt{2019MNRAS.484.4444U}). Thus, the column density ratio demonstrates a more distinct increasing trend than the integrated intensity ratio. The properties of these observed sources, including heliocentric distances, temperatures, masses, and the column densities of HC$_3$N and N$_2$H$^+$, have been provided by \citet{2019ApJ...872..154T}, \citet{2023ApJS..264...48W}, and relevant references, which exclude potential biases. However, some observed ratios between HMSC and HMPO stages or between HMPO and UCH{\sc ii} stages can overlap (see Fig. 10 from \citealt{2019ApJ...872..154T} and Fig. 6 from \citealt{2023ApJS..264...48W}), which implies that a particular source may not exhibit a strictly increasing ratio of HC$_3$N/N$_2$H$^+$ throughout the evolution. Besides, the observed HC$_3$N/N$_2$H$^+$ of different sources in each evolutionary stage can fluctuate by over one order of magnitude. This suggests that observed sources in the same stage may span a considerable duration of time, and the value of HC$_3$N/N$_2$H$^+$ during a certain period in the HMSC, HMPO, and UCH{\sc ii} stage could present an increasing trend. Thus, it is necessary to constrain the potential timescales of each stage by fitting the median value of observed HC$_3$N/N$_2$H$^+$ via chemical simulations. 
  
  \par{}\citet{2014A&A...563A..97G, 2015A&A...579A..80G} constrained the physical conditions and timescales of each stage in a 1D approximation by fitting the abundances of 18 species (including four deuterated species) toward 59 HMSFRs (19 IRDCs, 20 HMPOs, 11 HMCs, and 9 UCH{\sc ii} regions) through the IRAM 30 m, which can be considered as fiducial physical models in chemical simulations for further comparing other observed ratios during each evolutionary stage. Based on such simulations and observations, 15 column density ratios between different species (excluding deuterated species) were also compared. Only CS/SO, HCN/HNCO, HCO$^+$/N$_2$H$^+$, HNCO/C$_2$H, and N$_2$H$^+$/C$_2$H show monotonic evolutionary trends. However, none of these simulated ratios fit the observed ratios well; indeed, they even exhibit opposite trends, such as HCO$^+$/N$_2$H$^+$ and N$_2$H$^+$/C$_2$H. Additionally, such fiducial models had been further adopted for fitting observed abundances and estimating the lifetime of each HMSFR \citep[e.g.,][]{2021A&A...648A..66G, 2023MNRAS.518.1472Y}. \citet{2019ApJ...881...57T, 2021ApJ...908..100T, 2023ApJS..267....4T} adopted homogeneous one-point physical models that include a free-fall collapse phase, a warm-up phase, and a hot phase \citep{2006A&A...457..927G} to calculate the ratios between several complex organic molecules (COMs; HC$_5$N/CH$_3$OH, C$_2$H/HC$_5$N, HC$_5$N/CH$_3$CN, CH$_3$OH/CH$_3$CN, CH$_3$OCH$_3$/CH$_3$CN, HNCO/CH$_3$CN, CH$_2$CHCN/CH$_3$CN, and CH$_3$CH$_2$CN/CH$_3$CN) toward several MYSOs. These modeled ratios can constrain a hot core stage ($T>100\,\mathrm{K}$) but show no clear evolutionary trend, while the one-point physical models were not sufficient to accurately describe HMSFRs. Thus, a systematic theoretical search for chemical clocks via chemical modeling is necessary.
  
  \par{}In this paper, we perform chemical simulations to replicate the observed column density ratio of HC$_3$N/N$_2$H$^+$ and the abundance of these two species across various evolutionary stages of HMSFRs. We focus on this promising chemical clock candidate because it has not been explicitly considered as an evolutionary indicator and analyzed in previous simulations. Simultaneously, we identify the chemical processes responsible for the observed time-dependent trends in these stages. Besides, we analyze other candidate chemical clocks according to numerical simulations. In Sect.~\ref{section2}, we describe adopted chemical models and several physical models for simulating chemical evolution within the range of $10^5\, \mathrm{AU}$ ($\sim0.5\,\mathrm{pc}$) in HMSFRs. The 1D models presented by \citep{2014A&A...563A..97G, 2015A&A...579A..80G} show fluctuations in density and temperature across the entire evolution. To maintain the steady increase in density and temperature over time as predictions of the global hierarchical collapse scenario, we also adjust parameters such as density, temperature, and time spent in each evolutionary stage. In Sect.~\ref{section3}, we compare the modeled average ratio of HC$_3$N/N$_2$H$^+$ within different radii with observed ratios, and present the evolution of HC$_3$N/N$_2$H$^+$ at different locations. In Sect.~\ref{section4}, we discuss key mechanisms affecting the evolution of HC$_3$N/N$_2$H$^+$, and propose other candidate chemical clocks in HMSFRs according to our simulations. Finally, in Sect.~\ref{section5} we present our conclusions.

  \section{Model} \label{section2}
      \subsection{Chemical model} \label{section2.1}
 
   \par{}To compare the observations with theoretical predictions in astrochemistry, the evolutions of HC$_3$N and N$_2$H$^+$ were calculated via the astrochemical code Nautilus \citep{2016MNRAS.459.3756R}. In the simulations, both the two-phase (gas and grain surface) model \citep{1992ApJS...82..167H} and the three-phase (gas, grain surface, and grain mantle) model \citep{1993MNRAS.263..589H} were adopted. Since the results adopting the latter one deviate from the observed HC$_3$N/N$_2$H$^+$ because most species in the grain mantles are chemically inert when the temperature is low ($<50\, \mathrm{K}$), only the simulations using the two-phase model are discussed. The reaction network is identical to the one proposed by \citet{2021A&A...648A..72W} encompassing 9710 reactions and 947 species, which already includes the photodesorption mechanism \citep{2009A&A...496..281O, 2014ApJ...787..135C}. To exclude potential bugs, this reaction network was benchmarked against the reaction network from \citet{2016MNRAS.459.3756R} in the Nautilus package (see Appendix \ref{appendix_benchmark}),  which includes 717 species and over 11000 reactions. A higher cosmic ray ionization rate of $\zeta_{\mathrm{CR}} = 1.8\times10^{-16} \, \mathrm{s^{-1}}$ was utilized according to the survey of 20 HMSFRs in the Galactic disk performed by \citet{2015ApJ...800...40I}; according to these authors, $\zeta_{\mathrm{CR}}$ is supposed to remain constant at a Galactic radius $>5\,\mathrm{kpc}$. Given the distances of observed sources from \citet{2019ApJ...872..154T} and \citet{2023ApJS..264...48W}, it is reasonable to adopt this value, which is one order of magnitude higher than the common standard value of $1.3\times10^{-17} \, \mathrm{s^{-1}}$ for the ISM. This rate is assumed to remain constant throughout the evolution, following the chemical simulations conducted by \citet{2021A&A...648A..66G}. It should be noted that cosmic rays are attenuated inside molecular clouds as a function of column density \citep{2022A&A...658A.189P}. However, for the inner compact regions of HMSFRs with higher densities and temperatures, several chemical mechanisms become more significant compared to those in the outer diffuse areas. For example, the reaction rates of gaseous reactions and thermal desorption can be several orders of magnitude higher than those in the outer regions. This makes the cosmic ray ionization and dissociation less critical. We compared the results of HC$_3$N and N$_2$H$^+$, adopting $\zeta_{\mathrm{CR}} = 1.8\times10^{-16} \, \mathrm{s^{-1}}$ and $1.3\times10^{-17} \, \mathrm{s^{-1}}$, respectively, using the physical model proposed by \citet[see our Sect.~\ref{section2.2}]{2014A&A...563A..97G}. For the inner compact regions, the deviations between these two simulations throughout the entire evolution are fewer than several times. Therefore, for simplicity, a constant ionization rate across HMSFRs was adopted. Similarly, given the potentially stronger FUV radiation field in HMSFRs, $\chi$ was assumed to be 10 times higher than the standard value of 1, just like $\zeta_{\mathrm{CR}}$. Actually, considering the high value of visual extinction, $A_{\mathrm{V}}$ ($>10\, \mathrm{mag}$), in the majority of HMSFRs, the impact of such an enhanced radiation field on the results can be ignored. We also adopted the sticking coefficient of adsorption on the grains of $P_{\mathrm{s}}=1$, the reactive desorption efficiency of $a_{\mathrm{RRK}}=0.01$ \citep{2007A&A...467.1103G, 2013ApJ...769...34V}, the grain surface diffusion barrier of $E_{\mathrm{b1}}=0.5E_{\mathrm{d}}$, and the grain mantle diffusion barrier of $E_{\mathrm{b2}}=0.7E_{\mathrm{d}}$ \citep{2016ApJ...819..145C}, where $E_{\mathrm{d}}$ is the desorption energy of each species. Other parameters can be consulted in Table 1 presented by \citet{2019A&A...622A.185W}. In addition, the initial abundances listed in Table \ref{table1} approximate the low-metal environment at the start of the HMSC stage \citep{2014A&A...563A..97G}.

      \begin{table}
        \caption{Initial abundances \citep{2014A&A...563A..97G}. }
        \label{table1}
        \begin{center}
          \begin{tabular}{ll}\\
          \hline
          \hline
            Species&$n_{\mathrm{X}}/n_{\mathrm{H}}$\\
            \hline
            H$_{2}$&4.99(-1)\tablefootmark{a}\\
            H&2.00(-3)\\
            He&9.75(-2)\\
            C&7.86(-5)\\
            N&2.47(-5)\\
            O&1.80(-4)\\
            S&8.00(-7)\\
            Na&2.25(-9)\\
            Mg&1.09(-8)\\
            Si&3.00(-9)\\
            P&2.16(-10)\\
            Cl&1.00(-9)\\
            Fe&2.74(-9)\\
          \hline\\
          \end{tabular}\\
          \tablefoot{\tablefoottext{a}{$a(b)=a\times10^b .$}}
        \end{center}
      \end{table}

      \subsection{Physical model} \label{section2.2}

     \begin{table*}
        \caption{Physical models of HMSFRs. }
        \label{table2}
        \begin{center}
          \resizebox{\textwidth}{!}{
          \begin{tabular}{lllllll}\\
          \hline
          \hline
            Parameter&Model 1\tablefootmark{a}&Model 2\tablefootmark{b}&Model 3&Model 4&Model 5&Model 6\\
            \hline
            HMSC stage&&&&&&\\
            Inner radius $r_{\mathrm{in}}$&102 AU&12700 AU&102 AU&12700 AU&102 AU&102 AU\\
            Outer radius $r_{\mathrm{out}}$&$10^5 \, \mathrm{AU}$&$10^5 \, \mathrm{AU}$&$10^5 \, \mathrm{AU}$&$10^5 \, \mathrm{AU}$&$10^5 \, \mathrm{AU}$&$10^5 \, \mathrm{AU}$\\
            Density at the inner radius $\rho_{\mathrm{in}}$&$1.2\times10^9\, \mathrm{cm^{-3}}$&$1.4\times10^5\, \mathrm{cm^{-3}}$&$1.2\times10^9\, \mathrm{cm^{-3}}$&$1.4\times10^5\, \mathrm{cm^{-3}}$&$1.2\times10^9\, \mathrm{cm^{-3}}$&$1.2\times10^9\, \mathrm{cm^{-3}}$\\
            Density power-law index $p$&1.8&1.5&1.8&1.5&1.8&1.8\\
            Temperature at the inner radius $T_{\mathrm{in}}$&20.9 K&11.3 K&20.9 K&11.3 K&20.9 K&20.9 K\\
            Temperature power-law index $q$&0&0&0&0&0&0\\
            Timescale of HMSC stage $t_{\mathrm{HMSC}}$&$1.1\times10^4\, \mathrm{yr}$&$1.65\times10^4\, \mathrm{yr}$&$3.0\times10^4\, \mathrm{yr}$&$3.0\times10^4\, \mathrm{yr}$&$3.0\times10^4\, \mathrm{yr}$&$3.0\times10^4\, \mathrm{yr}$\\
            \hline
            HMPO stage&&&&&&\\
            Inner radius $r_{\mathrm{in}}$&1130 AU&103 AU&1130 AU&103 AU&102 AU&102 AU\\
            Outer radius $r_{\mathrm{out}}$&$10^5 \, \mathrm{AU}$&$10^5 \, \mathrm{AU}$&$10^5 \, \mathrm{AU}$&$10^5 \, \mathrm{AU}$&$10^5 \, \mathrm{AU}$&$10^5 \, \mathrm{AU}$\\
            Density at the inner radius $\rho_{\mathrm{in}}$&$5.0\times10^6\, \mathrm{cm^{-3}}$&$1.5\times10^9\, \mathrm{cm^{-3}}$&$5.0\times10^6\, \mathrm{cm^{-3}}$&$1.5\times10^9\, \mathrm{cm^{-3}}$&$1.5\times10^9\, \mathrm{cm^{-3}}$&$1.5\times10^9\, \mathrm{cm^{-3}}$\\
            Density power-law index $p$&1.5&1.8&1.5&1.8&1.8&1.8\\
            Temperature at the inner radius $T_{\mathrm{in}}$&77.3 K&75.8 K&77.3 K&75.8 K&227.4 K&227.4 K\\
            Temperature power-law index $q$&0.4&0.4&0.4&0.4&0.4&0.4\\
            Timescale of HMPO stage $t_{\mathrm{HMPO}}$&$6.0\times10^4\, \mathrm{yr}$&$3.2\times10^4\, \mathrm{yr}$&$3.0\times10^4\, \mathrm{yr}$&$3.0\times10^4\, \mathrm{yr}$&$3.0\times10^4\, \mathrm{yr}$&$3.0\times10^4\, \mathrm{yr}$\\
            \hline
            HMC stage&&&&&&\\
            Inner radius $r_{\mathrm{in}}$&102 AU&1140 AU&102 AU&1140 AU&102 AU&102 AU\\
            Outer radius $r_{\mathrm{out}}$&$10^5 \, \mathrm{AU}$&$10^5 \, \mathrm{AU}$&$10^5 \, \mathrm{AU}$&$10^5 \, \mathrm{AU}$&$10^5 \, \mathrm{AU}$&$10^5 \, \mathrm{AU}$\\
            Density at the inner radius $\rho_{\mathrm{in}}$&$2.7\times10^9\, \mathrm{cm^{-3}}$&$1.3\times10^8\, \mathrm{cm^{-3}}$&$2.7\times10^9\, \mathrm{cm^{-3}}$&$1.3\times10^8\, \mathrm{cm^{-3}}$&$2.7\times10^9\, \mathrm{cm^{-3}}$&$2.7\times10^9\, \mathrm{cm^{-3}}$\\
            Density power-law index $p$&1.9&2.0&1.9&2.0&1.9&1.6\\
            Temperature at the inner radius $T_{\mathrm{in}}$&268.3 K&162.9 K&268.3 K&162.9 K&268.3 K&268.3 K\\
            Temperature power-law index $q$&0.4&0.4&0.4&0.4&0.4&0.34\\
            Timescale of HMC stage $t_{\mathrm{HMC}}$&$4.2\times10^4\, \mathrm{yr}$&$3.5\times10^4\, \mathrm{yr}$&$3.0\times10^4\, \mathrm{yr}$&$3.0\times10^4\, \mathrm{yr}$&$3.0\times10^4\, \mathrm{yr}$&$3.0\times10^4\, \mathrm{yr}$\\
            \hline
            UCH{\sc ii} stage&&&&&&\\
            Inner radius $r_{\mathrm{in}}$&102 AU&103 AU&102 AU&103 AU&102 AU&102 AU\\
            Outer radius $r_{\mathrm{out}}$&$10^5 \, \mathrm{AU}$&$10^5 \, \mathrm{AU}$&$10^5 \, \mathrm{AU}$&$10^5 \, \mathrm{AU}$&$10^5 \, \mathrm{AU}$&$10^5 \, \mathrm{AU}$\\
            Density at the inner radius $\rho_{\mathrm{in}}$&$2.4\times10^8\, \mathrm{cm^{-3}}$&$1.0\times10^{10}\, \mathrm{cm^{-3}}$&$2.4\times10^8\, \mathrm{cm^{-3}}$&$1.0\times10^{10}\, \mathrm{cm^{-3}}$&$1.0\times10^{10}\, \mathrm{cm^{-3}}$&$1.0\times10^{10}\, \mathrm{cm^{-3}}$\\
            Density power-law index $p$&1.5&2.0&1.5&2.0&2.0&1.6\\
            Temperature at the inner radius $T_{\mathrm{in}}$&293.1 K&244.3 K&293.1 K&244.3 K&293.1 K&293.1 K\\
            Temperature power-law index $q$&0.4&0.4&0.4&0.4&0.4&0.34\\
            Timescale of UCH{\sc ii} stage $t_{\mathrm{UCHII}}$&$1.2\times10^4\, \mathrm{yr}$&$1.65\times10^4\, \mathrm{yr}$&$3.0\times10^4\, \mathrm{yr}$&$3.0\times10^4\, \mathrm{yr}$&$3.0\times10^4\, \mathrm{yr}$&$3.0\times10^4\, \mathrm{yr}$\\
          \hline
          Total timescale $t_{\mathrm{total}}$&$1.25\times10^5\, \mathrm{yr}$&$1.0\times10^5\, \mathrm{yr}$&$1.2\times10^5\, \mathrm{yr}$&$1.2\times10^5\, \mathrm{yr}$&$1.2\times10^5\, \mathrm{yr}$&$1.2\times10^5\, \mathrm{yr}$\\
          \hline\\
          \end{tabular}}\\
          \tablefoot{\tablefoottext{a}{\citet{2014A&A...563A..97G}. }\tablefoottext{b}{\citet{2015A&A...579A..80G}. } }
        \end{center}
      \end{table*}

    \par{}We adopted the 1D physical model of HMSFRs including four evolutionary stages derived by \citet{2014A&A...563A..97G, 2015A&A...579A..80G}, and their parameters are presented in Table \ref{table2}, designated as model 1 and model 2, respectively, serving as benchmark physical models. The densities and the temperatures at different radii were calculated with modified power laws,  
         \begin{align}
          \rho(r)=\rho_{\mathrm{in}}(r/r_{\mathrm{in}})^{-p}, \, r \ge r_{\mathrm{in}}; \label{equation1} \\
          \rho(r)=\rho_{\mathrm{in}}, \, r < r_{\mathrm{in}};
          \end{align}     
and
         \begin{align}
          T(r)=T_{\mathrm{in}}(r/r_{\mathrm{in}})^{-q}, \, r \ge r_{\mathrm{in}}; \label{equation2} \\
          T(r)=T_{\mathrm{in}}, \, r < r_{\mathrm{in}}, 
          \end{align}
which remain constant during each stage. For simplicity, the gas temperature and the grain temperature were assumed to be the same. The UCH{\sc ii} lifetime in model 2 was set to be $1.65\times10^4\, \mathrm{yr}$ instead of $3.0\times10^3\, \mathrm{yr}$ for consistency with model 1 and an easy analysis with $t_{\mathrm{total}}=1.0\times10^5\, \mathrm{yr}$. However, simulations adopting these two models cannot fit the observed evolutionary trend of HC$_3$N/N$_2$H$^+$. To improve the fit between simulations and observations without altering the density distribution or the temperature distribution, the timescales of each stage were adjusted considering an uncertainty of a factor of $2-3$ for the best-fit ages proposed by \citet{2014A&A...563A..97G, 2015A&A...579A..80G}. Consequently, models 3 and 4 only modify the duration of each stage to be $3\times10^4 \, \mathrm{yr}$ compared to model 1 and 2, respectively, with prolonged timescales for the HMSC and UCH{\sc ii} stage, but slightly reduced timescales for the HMPO and HMC stage. Such HMSC and HMPO timescales are consistent with the statistical lifetimes for clumps with masses of $\sim6.3\times10^3\,M_{\sun}$ \citep[see Table 6 from][]{2018MNRAS.473.1059U}, while HMC and UCH{\sc ii} timescales are shorter by a factor of $3-4$. Nevertheless, according to the simulated evolutionary trend of HC$_3$N/N$_2$H$^+$ in the HMC and UCH{\sc ii} stage, if we adopt longer HMC and UCH{\sc ii} timescales (e.g., $10^5\,\mathrm{yr}$), the ratio could be much lower than observed values in UCH{\sc ii} regions. Therefore, the suggested durations of the HMC and UCH{\sc ii} stage remain reasonable nonetheless. Additionally, $\rho_{\mathrm{in}}$ does not consistently increase throughout the evolution accompanied by the variations in $r_{\mathrm{in}}$ in both models 1 and 2. To maintain the same $r_{\mathrm{in}}$ with reasonable increasing $\rho_{\mathrm{in}}$ and $T_{\mathrm{in}}$ throughout the entire evolution, model 5 combines the parameters from the HMSC and HMC stage of model 3 with those from the HMPO and UCH{\sc ii} stage of model 4. A higher $T_{\mathrm{in}}$ in the HMPO stage was adopted to achieve a better agreement with observations. Finally, to have a consistently increased density and temperature within the outer regions ($r\sim r_{\mathrm{out}}$) during the late stages, model 6 adopts reduced values of $p$ and $q$ in the HMC and UCH{\sc ii} stage compared to model 5. Similarly, other test models adopting different values of $p$ ($1.5-2.0$) and $q$ ($0.3-0.4$) based on model 5 were calculated but are no longer discussed in this paper, since these results cannot fit the observations well. The visual extinction was calculated by using $A_{\mathrm{V}}(r)=A_{\mathrm{V_0}}(n_{\mathrm{H}}(r)/n_{\mathrm{H_0}})^{2/3}$ with $A_{\mathrm{V_0}}=5\, \mathrm{mag}$ and $n_{\mathrm{H_0}}=10^4\, \mathrm{cm^{-3}}$ \citep{2019ApJ...881...57T}, while $n_{\mathrm{H}}(r)=2\rho(r)$. For simplicity, the physical parameters undergo an instantaneous variation at the transition point between two stages, as was proposed by \citet{2014A&A...563A..97G, 2015A&A...579A..80G}. For each physical model, we selected about 45 different locations from the center to the outmost radius of $10^5 \, \mathrm{AU}$ ($\sim0.5\,\mathrm{pc}$) for chemical modeling, then the abundances at different radii and the average abundances within different ranges in an HMSFR could be computed. The average number density of each species is equal to the summed molecule number in each shell divided by the total volume of the same shell; at this point, the average abundance with respect to H$_2$ can be calculated.

  \section{Results} \label{section3}
      
      \subsection{Comparison of chemical simulations and observations} \label{section3.1}

 \begin{figure*}
 \centering
 \includegraphics[width=1.0\textwidth]{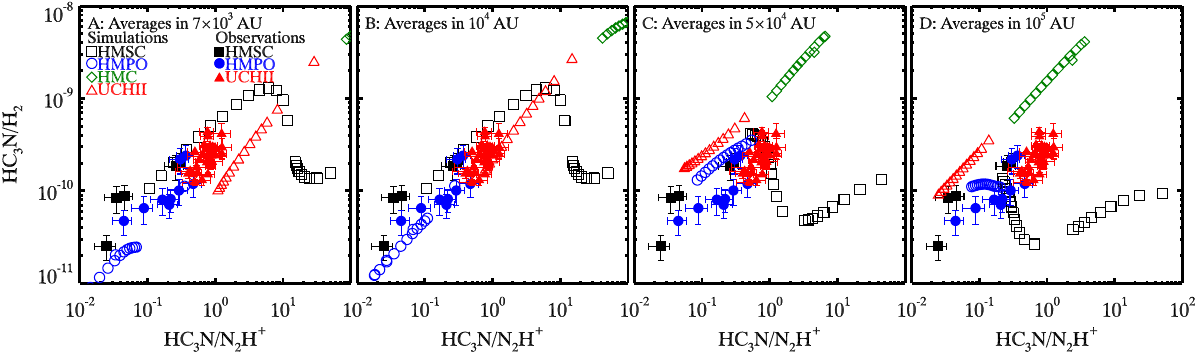}
     \caption{Correlation between the average ratio of HC$_{3}$N/N$_{2}$H$^{+}$ and the average abundance of HC$_{3}$N with respect to H$_2$ within four different ranges (panel A: $7\times10^3\, \mathrm{AU}$, panel B: $10^4\, \mathrm{AU}$, panel C: $5\times10^4\, \mathrm{AU}$, and panel D: $10^5\, \mathrm{AU}$) around the center by adopting model 4 in chemical simulations compared with observations. The results during the HMSC, HMPO, HMC, and UCH{\sc ii} stages are plotted as squares, circles, diamonds, and triangles, respectively. The evolution of HC$_{3}$N/N$_{2}$H$^{+}$ and HC$_{3}$N/H$_2$ throughout $t_{\mathrm{total}}$ during each stage consistently decrease (from right to left and from top to bottom). The observations proposed by \citet{2019ApJ...872..154T} and \citet[including derived column density of HC$_{3}$N, N$_{2}$H$^{+}$, and H$_2$ simultaneously]{2023ApJS..264...48W}  are plotted as solid squares, solid circles, and solid triangles, respectively, with error bars.}
     \label{figure1}
 \end{figure*}

   \begin{figure*}
 \centering
 \includegraphics[width=1.0\textwidth]{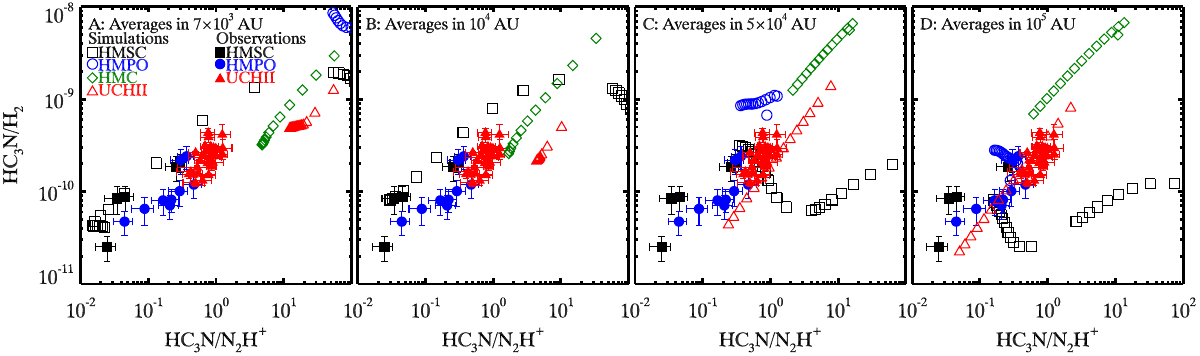}
     \caption{Same as Fig. \ref{figure1}, but adopting model 6.}
     \label{figure2}
 \end{figure*}

 \par{}The sources observed by \citet{2019ApJ...872..154T} and \citet{2023ApJS..264...48W} are mainly located at a distance of $\sim3-10\, \mathrm{kpc}$. Considering the largest $29\arcsec$ beam size of the IRAM 30 m, the linear resolution corresponds to $\sim0.4-1.4\, \mathrm{pc}$. Since $r_{\mathrm{out}}=10^5 \, \mathrm{AU}$ is comparable to the linear resolution, and also approximately corresponds to the outermost radius of dense parts of HMSFRs \citep{2014A&A...563A..97G}, the average abundances of HC$_3$N and N$_2$H$^+$ with respect to H$_2$ within a radius of $10^5\, \mathrm{AU}$ were calculated, and the corresponding ratio HC$_3$N/N$_2$H$^+$ was compared with observed targets \citep{2019ApJ...872..154T, 2023ApJS..264...48W}. To investigate the correlation between the simulated average ratio of HC$_{3}$N/N$_{2}$H$^{+}$ and HC$_{3}$N/H$_2$ (the abundance of HC$_{3}$N with respect to H$_2$) and compare them with observations, we plot them as scatter diagrams in Figs. \ref{figure1} and \ref{figure2}, within four different ranges ($7\times10^3\, \mathrm{AU}$, $10^4\, \mathrm{AU}$, $5\times10^4\, \mathrm{AU}$, and $10^5\, \mathrm{AU}$) around the center using models 4 and 6, respectively. These figures show that there is only one radius per model ($10^4\, \mathrm{AU}$ for model 4 and $10^5\, \mathrm{AU}$ for model 6) that can reproduce three of the four stages. The detailed comparison between models and observations is discussed below. We note already that due to a lack of observational constraints and observational ambiguities, we did not consider the HMC stage for comparison with our models. Such different ranges were chosen to verify whether the average ratios within certain ranges can fit the observations. The average ratios within $7\times10^3\, \mathrm{AU}$ usually severely deviate from the observations. The evolution during the HMSC stage, HMPO stage, HMC stage, and UCH{\sc ii} stage is depicted with different symbols, indicating that the correlation consistently evolves from right to left with a decreasing HC$_3$N/N$_2$H$^+$ ratio, while it moves from top to bottom with a decreasing HC$_3$N/H$_2$ ratio in each panel. A line depicting the evolutionary trend is omitted in each panel for concise pictures, since abrupt jumps are induced by discontinuous changes in the physical parameters between the two stages. However, such abrupt jumps do not affect the comparison between modeled and observed ratios. The average ratios within $10^4\, \mathrm{AU}$ around the center can fit the observations at the final HMSC stage ($t_{\mathrm{total}}\sim2.8-3.0\times10^4\, \mathrm{yr}$), the start HMPO stage ($t_{\mathrm{total}}\sim3.0-3.1\times10^4\, \mathrm{yr}$), and the final UCH{\sc ii} stage ($t_{\mathrm{total}}\sim1.03-1.2\times10^5\, \mathrm{yr}$), respectively (see panel B in Fig. \ref{figure1}). Nevertheless, on the one hand, it is noteworthy that a strictly monotonic increase in HC$_3$N/N$_2$H$^+$ over the entire evolution cannot be achieved. On the other hand, such a region only accounts for 10\% of the radius and $10^{-3}$ of the volume compared to the entire area of HMSFRs, which seems unlikely to reflect the actual conditions. The simulations within the other three ranges cannot fit the observations, with a lower HC$_3$N/H$_2$ ratio in the HMPO stage and a higher HC$_{3}$N/N$_2$H$^+$ ratio in the UCH{\sc ii} stage within the range of $7\times10^3\, \mathrm{AU}$ (panel A), a higher HC$_{3}$N/N$_2$H$^+$ ratio in the HMSC stage and a lower HC$_{3}$N/N$_2$H$^+$ ratio in the UCH{\sc ii} stage within the range of $5\times10^4\, \mathrm{AU}$ (panel C), and a lower HC$_{3}$N/N$_2$H$^+$ ratio in the UCH{\sc ii} stage within the range of $10^5\, \mathrm{AU}$ (panel D). Additionally, for the inner region within $7\times10^3\, \mathrm{AU}$, both simulated HC$_{3}$N/N$_{2}$H$^{+}$ and HC$_{3}$N/H$_2$ ratios can deviate from the observations by over one order of magnitude, because of the high temperature and density. These two simulated ratios can only fit the observations at the late HMSC stage. Thus, such results are no longer plotted.

 \par{}Similarly, Fig. \ref{figure2} depicts the comparison between the simulations using model 6 and observations. In panel D, the average ratios within $10^5\, \mathrm{AU}$ from the center can fit observations at the late HMSC stage ($t_{\mathrm{total}}\sim2.0-3.0\times10^4\, \mathrm{yr}$), the early HMPO stage ($t_{\mathrm{total}}\sim3.0-3.8\times10^4\, \mathrm{yr}$), and the early UCH{\sc ii} stage ($t_{\mathrm{total}}\sim9.4-10.1\times10^4\, \mathrm{yr}$), respectively. Therefore, this simulation is more reasonable than the result using model 4, although the strictly monotonic increase in HC$_3$N/N$_2$H$^+$ throughout the entire evolution remains unreplicated. The results within more internal regions deviate from observations more significantly, especially during the HMPO and UCH{\sc ii} stage (see panels A, B, and C), due to the influence of much higher temperature and density in the innermost regions. In addition, the results adopting the other four models are presented in Appendix \ref{appendixA}, which cannot fit the observations regardless of the radius of the selected regions. Consequently, we propose that the observed increased HC$_3$N/N$_2$H$^+$ ratio from the HMSC stage to the HMPO stage and to the UCH{\sc ii} stage in HMSFRs can be fit during a certain period in each stage in chemical simulations, which may constrain the evolution at the late HMSC stage, the early HMPO stage, and the early UCH{\sc ii} stage.

 \par
\begin{figure*}
 \centering
  \includegraphics[width=1.0\textwidth]{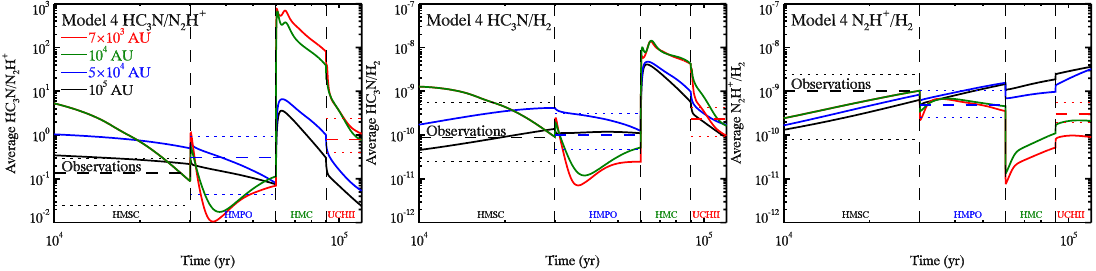}
  \includegraphics[width=1.0\textwidth]{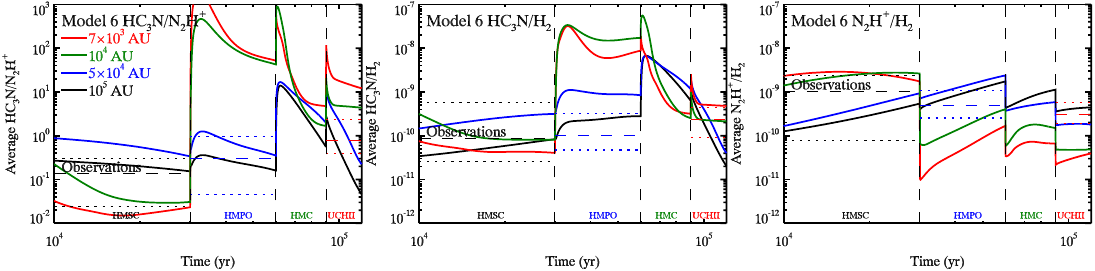}
     \caption{Time-dependent evolution of the average ratio of HC$_{3}$N/N$_{2}$H$^{+}$ and the average fractional abundance of HC$_{3}$N and N$_{2}$H$^{+}$ with respect to H$_2$ within four different ranges ($7\times10^3\, \mathrm{AU}$, $10^4\, \mathrm{AU}$, $5\times10^4\, \mathrm{AU}$, and $10^5\, \mathrm{AU}$) around the center using model 4 (left panels) and model 6 (right panels). The four evolutionary stages are labeled in each panel, and the observations are also plotted, with the dashed lines as the median values and the dotted lines as the maximums or minimums of observations. The evolution before $10^4\, \mathrm{yr}$ is excluded to show the changes in different stages more clearly. The results within $7\times10^3\, \mathrm{AU}$ and $10^4\, \mathrm{AU}$ during the HMSC stage in model 4 completely overlap because of the large $r_{\mathrm{in}}>10^4\,\mathrm{AU}$. } 
     \label{figure3}
 \end{figure*}

{}To individually analyze the time-dependent evolution of average ratios of HC$_{3}$N/N$_{2}$H$^{+}$, HC$_{3}$N/H$_{2}$, and N$_{2}$H$^{+}$/H$_2$ in models 4 and 6, we depict the average ratios within the four different aforementioned ranges ($7\times10^3\, \mathrm{AU}$, $10^4\, \mathrm{AU}$, $5\times10^4\, \mathrm{AU}$, and $10^5\, \mathrm{AU}$) around the center across $t_{\mathrm{total}}$ in Fig. \ref{figure3}. In model 4, the average ratio of HC$_{3}$N/N$_{2}$H$^{+}$ decreases during the HMSC stage and fits the observations at the final HMSC stage, particularly the ratio within the region of $10^4\, \mathrm{AU}$. Then an abrupt increase occurs at the start HMPO stage within the innermost $10^4\, \mathrm{AU}$ region, mainly induced by the desorption of HC$_3$N on the grains due to high temperatures. However, the ratio rapidly decreases to be lower than the observations in this region, while within larger regions ($5\times10^4\, \mathrm{AU}$ and $10^5\, \mathrm{AU}$) the ratio continues to decrease from the HMSC stage. During the HMC stage, the ratio similarly increases at the start because of high temperatures but subsequently decreases within all four regions. The ratio continues to decrease during the UCH{\sc ii} stage, but the values within the innermost $10^4\, \mathrm{AU}$ region can fit the observations, while those within larger regions ($5\times10^4\, \mathrm{AU}$ and $10^5\, \mathrm{AU}$) can be lower than the observations. For the average abundance of HC$_{3}$N, the evolutionary tendency is similar to that of HC$_{3}$N/N$_{2}$H$^{+}$, but the results within all four regions can roughly agree with the observations. The average abundance of N$_{2}$H$^{+}$ increases during each stage across all four regions \citep[also see][]{2014A&A...563A..97G, 2015A&A...579A..80G} in contrast to the observed decreased abundance of N$_{2}$H$^{+}$, but only the result within $10^4\, \mathrm{AU}$ can fit the observations. In model 6, the average ratio of HC$_{3}$N/N$_{2}$H$^{+}$ still exhibits a decreasing trend during each stage, with sharp increases at the start of each stage compared with the values at the end of the previous stage. The ratio within $7\times10^3\, \mathrm{AU}$ or $10^4\, \mathrm{AU}$ is lower than the observations during the HMSC stage, but much higher than the observations during the HMPO and UCH{\sc ii} stage. The ratio within $5\times10^4\, \mathrm{AU}$ is higher than the observations during the HMSC stage, but can fit the observations during the HMPO and UCH{\sc ii} stage. Only the ratio within $10^5\, \mathrm{AU}$ can fit the observations during all three stages. For the average ratio of HC$_{3}$N/H$_2$, the results within four regions can agree with the observations during the HMSC and UCH{\sc ii} stage, but only the result within $10^5\, \mathrm{AU}$ can fit the observations during the HMPO stage. For the average ratio of N$_{2}$H$^+$/H$_2$, the tendency to increase occurs during each stage accompanied by a dramatic decrease at the start of each stage compared to the values at the end of the previous stage. Still, only the result within $10^5\, \mathrm{AU}$ can fit the observations during the UCH{\sc ii} stage. Obviously, a rigorously monotonically increasing ratio of HC$_{3}$N/N$_{2}$H$^{+}$ throughout the evolution of HMSFRs cannot be accurately replicated in simulations.

       \begin{table*}
        \caption{Observations and best-fit simulations of HC$_{3}$N/N$_{2}$H$^{+}$, HC$_{3}$N/H$_{2}$, and N$_{2}$H$^{+}$/H$_2$ in HMSFRs. }
        \label{table3}
        \begin{center}
        \resizebox{\textwidth}{!}{
          \begin{tabular}{lcccc}\\
          \hline
          \hline
            Ratio&Observation&Median value\tablefootmark{a}&Within $10^4\,\mathrm{AU}$ in model 4&Within $10^5\,\mathrm{AU}$ in model 6\\
            \hline
            HMSC stage&&&&\\
            HC$_{3}$N/N$_{2}$H$^{+}$&$0.0247-0.298$&0.136&0.135&0.154\\
            HC$_{3}$N/H$_{2}$&$2.53\times10^{-11}-5.61\times10^{-10}$&$8.74\times10^{-11}$&$1.28\times10^{-10}$&$8.35\times10^{-11}$\\
            N$_{2}$H$^{+}$/H$_2$&$7.79\times10^{-11}-2.44\times10^{-9}$&$1.02\times10^{-9}$&$9.47\times10^{-10}$&$5.43\times10^{-10}$\\
            $\kappa$&-&-&0.947&0.908\\
            $t_{\mathrm{total}}$ (yr)&-&-&$2.78\times10^4$&$3.00\times10^4$\\
            \hline
            HMPO stage&&&&\\
            HC$_{3}$N/N$_{2}$H$^{+}$&$0.0446-0.941$&0.303&0.301&0.298\\
            HC$_{3}$N/H$_{2}$&$4.73\times10^{-11}-3.15\times10^{-10}$&$1.01\times10^{-10}$&$1.21\times10^{-10}$&$1.42\times10^{-10}$\\
            N$_{2}$H$^{+}$/H$_2$&$2.54\times10^{-10}-1.06\times10^{-9}$&$4.86\times10^{-10}$&$4.01\times10^{-10}$&$4.78\times10^{-10}$\\
            $\kappa$&-&-&0.956&0.957\\
            $t_{\mathrm{total}}$ (yr)&-&-&$3.09\times10^4$&$3.12\times10^4$\\
            \hline
            UCH{\sc ii} stage&&&&\\
            HC$_{3}$N/N$_{2}$H$^{+}$&0.398-2.41&0.777&1.08&0.775\\
            HC$_{3}$N/H$_{2}$&$9.37\times10^{-11}-4.32\times10^{-10}$&$2.34\times10^{-10}$&$2.33\times10^{-10}$&$3.04\times10^{-10}$\\
            N$_{2}$H$^{+}$/H$_2$&$1.94\times10^{-10}-5.62\times10^{-10}$&$3.02\times10^{-10}$&$2.16\times10^{-10}$&$3.92\times10^{-10}$\\
            $\kappa$&-&-&0.923&0.940\\
            $t_{\mathrm{total}}$ (yr)&-&-&$1.12\times10^5$&$9.67\times10^4$\\
            \hline\\
          \end{tabular}}\\
          \tablefoot{\tablefoottext{a}{The median values are derived from observations proposed by \citet{2019ApJ...872..154T} and \citet{2023ApJS..264...48W}. }}
        \end{center}
      \end{table*}

 \par{}Due to the current lack of observations of HC$_3$N/N$_2$H$^+$ in HMCs, the evolution of HC$_3$N/N$_2$H$^+$ in chemical simulations is temporarily not regarded as a primary focus in the analysis. However, there are UCH{\sc ii} regions that have the characteristic spectrum of a hot core, and HMPOs can appear as hot cores, depending on the resolution and sensitivity in observations. These observational uncertainties affect the calculation of column densities and abundances, as well as the clump classification. In fact, a scatter of around one dex in the observed ratios appears within any evolutionary stage, which may induce the ratios during the HMC stage to overlap with those in the HMPO and UCH{\sc ii} stages. Additionally, this ratio during the HMC stage indeed follows the increasing trend, considering the integrated intensity ratio of HC$_3$N/N$_2$H$^+$ derived by \citet{2019MNRAS.484.4444U}. Consequently, it should be higher than that during the HMPO stage and lower than that during the UCH{\sc ii} stage. In the simulations mentioned above, both HC$_3$N/N$_2$H$^+$ and HC$_3$N/H$_2$ in the HMC stage reach higher values compared to those not only in the HMPO stage but also in the UCH{\sc ii} stage, which contradicts the observations. Such a discrepancy is mainly driven by increased temperatures in the larger areas among HMSFRs during the HMC stage, which induces the thermal desorption of HC$_3$N molecules from the grains and subsequently a rapid increase in HC$_3$N/N$_2$H$^+$ (see Sect.~\ref{section4.1}). Nevertheless, the average HC$_3$N/N$_2$H$^+$ ratio within $10^5\, \mathrm{AU}$ at the late HMC stage already approaches the observed values in UCH{\sc ii} regions (see Fig. \ref{figure3}). Combined with the decreasing trend of HC$_3$N/N$_2$H$^+$ in this stage, adopting a longer duration of this stage ($>4\times10^4\, \mathrm{yr}$) potentially makes the simulated ratio fit the observed values. In addition, varying temperature distributions across the radius during this stage in physical models could significantly affect the degree of the increase in HC$_3$N/N$_2$H$^+$, as the scale of the area where the thermal desorption of HC$_3$N occurs is directly influenced. Such a comparison could be further explored in a future study, when more observations and simulations are available.

   \par{}To further constrain the evolutionary timescale in each stage, we calculated the confidence level, $\kappa$, of HC$_{3}$N/N$_{2}$H$^{+}$, HC$_{3}$N/H$_2$, and N$_{2}$H$^{+}$/H$_2$ by adopting the ``mean confidence level'' method \citep[e.g., ][]{2007A&A...467.1103G, 2008ApJ...681.1385H, 2011ApJ...743..182H, 2019A&A...622A.185W, 2021A&A...648A..72W}. For one specific ratio, $i$, the agreement between the calculated value, $X_i$, and the observed value, $X_{\mathrm{obs,}i}$, is defined as the confidence level, $\kappa_i$, which is calculated by
        \begin{align}
        \kappa_i=\mathrm{erfc} \left( \frac{|\lg X_i - \lg X_{\mathrm{obs,}i}|}{\sqrt{2}\sigma} \right),  \label{equation3}
        \end{align}
with the complementary error function, $\mathrm{erfc}=1-\mathrm{erf}$, and the standard deviation, $\sigma=1$. A higher value of $\kappa_i$ ($0-1$) indicates a better agreement between the calculated and observed values, while $\kappa_i=0.317$ corresponds to the calculated value being one order of magnitude higher or lower than the observed value. Subsequently, the average $\kappa$ across three ratios, HC$_{3}$N/N$_{2}$H$^{+}$, HC$_{3}$N/H$_2$, and N$_{2}$H$^{+}$/H$_2$, during each stage was calculated. By identifying the maximum average $\kappa$, the best-fit timescale in each stage can be obtained, which can further constrain the corresponding physical parameters.

   \par{}Table \ref{table3} presents the best-fit timescales for the HMSC stage, HMPO stage, and UCH{\sc ii} stage within $10^4\, \mathrm{AU}$ in model 4 and within $10^5\, \mathrm{AU}$ in model 6, respectively, which can fit all three observed ratios. It is crucial to compare these timescales with previous theoretical studies such as \citet[models 1 and 2]{2014A&A...563A..97G, 2015A&A...579A..80G}, which can evaluate the reliability of these timescales and other physical parameters adopted in our modified models. In the HMSC stage, the best fit occurs at $t_{\mathrm{total}}\sim3\times10^4\, \mathrm{yr}$, which may imply that a lifetime, $t_{\mathrm{HMSC}}=3\times10^4\, \mathrm{yr}$, a factor of $2-3$ longer than models 1 and 2 \citep{2014A&A...563A..97G, 2015A&A...579A..80G} is required to decrease the ratio of HC$_{3}$N/N$_{2}$H$^{+}$. A quick evolution within about $10^3\, \mathrm{yr}$ in the HMPO stage ($t_{\mathrm{total}}\sim3.1\times10^4\, \mathrm{yr}$) can fit the observed ratios, which may signify that $t_{\mathrm{HMPO}}=3.0\times10^4\, \mathrm{yr}$ is sufficient compared to the longer timescale of $6.0\times10^4\, \mathrm{yr}$ in model 1 and the comparable timescale of $3.2\times10^4\, \mathrm{yr}$ in model 2. Finally, in the UCH{\sc ii} stage, a duration of about $6.7\times10^3\, \mathrm{yr}$ ($t_{\mathrm{total}}\sim9.67\times10^4\, \mathrm{yr}$) achieves the best fit, which approaches half of $t_{\mathrm{UCHII}}=1.2\times10^4\, \mathrm{yr}$ in model 1 and $t_{\mathrm{UCHII}}=1.65\times10^4\, \mathrm{yr}$ in model 2. Overall, both the best-fit timescale and the total duration of each stage in our models are compatible with those in models 1 and 2 with an uncertainty of a factor of $2-3$. The best-fit timescales indicate the order of magnitude of the duration of the process of high-mass star formation, as the modeled chemical composition at these timescales is in broad agreement with statistical estimates. In addition, due to the relatively stable chemical evolutionary trend, certain periods can be confirmed. Different durations of the previous stage mainly influence the abundances at the start of the subsequent stage, with fluctuations by a factor of several times, but do not significantly affect the overall chemical evolutionary trend. Such differences could affect the agreement between simulations and observations at the start of subsequent stages. Overall, our modified physical models incorporating higher densities and temperatures in the outer regions still remain reasonable and provide valuable insights into the evolutionary timescales of HMSFRs, since different HMSFRs could evolve with different physical parameters.

     \subsection{Evolution of HC$_3$N/N$_2$H$^+$ at different locations} \label{section3.2}

   \begin{figure*}
 \centering
  \includegraphics[width=1.0\textwidth]{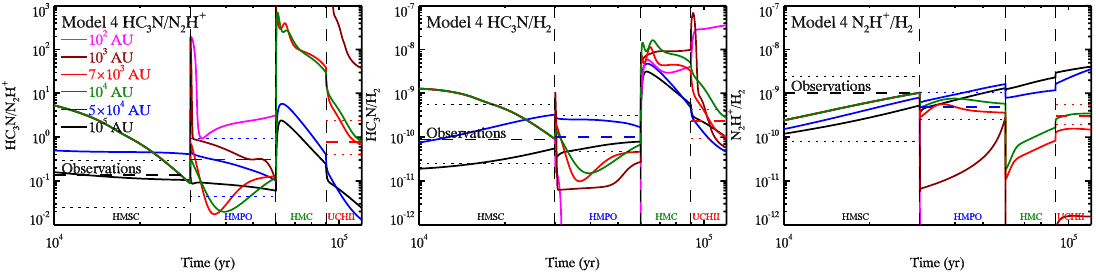}
  \includegraphics[width=1.0\textwidth]{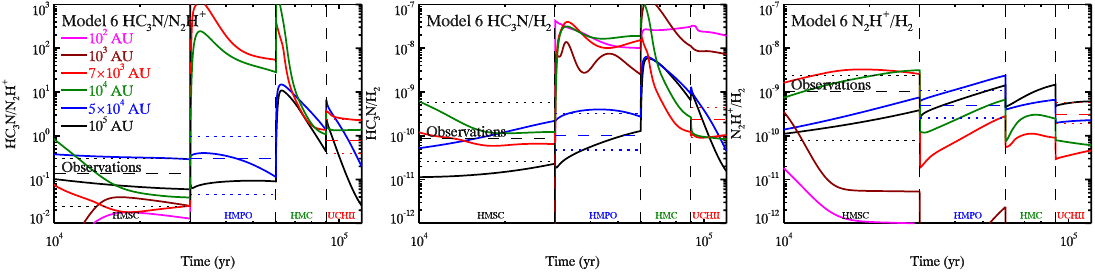}
     \caption{Same as Fig. \ref{figure3}, but depicting the ratio of HC$_{3}$N/N$_{2}$H$^{+}$ and the fractional abundance of HC$_{3}$N and N$_{2}$H$^{+}$ with respect to H$_2$ at six different radii ($10^2\, \mathrm{AU}$, $10^3\, \mathrm{AU}$, $7\times10^3\, \mathrm{AU}$, $10^4\, \mathrm{AU}$, $5\times10^4\, \mathrm{AU}$, and $10^5\, \mathrm{AU}$) from the center using model 4 (left panels) and model 6 (right panels). The results at $r=10^2\, \mathrm{AU}$, $10^3\, \mathrm{AU}$, $7\times10^3\, \mathrm{AU}$, and $10^4\, \mathrm{AU}$ during the HMSC stage in model 4 completely overlap because of the large $r_{\mathrm{in}}>10^4\,\mathrm{AU}$. } 
     \label{figure4}
 \end{figure*}

   \par{}To investigate potential physical parameters that mainly dominate the evolution of HC$_{3}$N/N$_{2}$H$^{+}$, it is necessary to analyze the agreement between observations and simulated ratios across different layers, spanning from the center to the outermost radius, $10^5\, \mathrm{AU}$, in the best-fit models. Figure \ref{figure4} presents the time-dependent evolution of HC$_{3}$N/N$_{2}$H$^{+}$, HC$_{3}$N/H$_{2}$, and N$_{2}$H$^{+}$/H$_2$ at six different radii ($10^2\, \mathrm{AU}$, $10^3\, \mathrm{AU}$, $7\times10^3\, \mathrm{AU}$, $10^4\, \mathrm{AU}$, $5\times10^4\, \mathrm{AU}$, and $10^5\, \mathrm{AU}$) from the center using model 4 and model 6. The evolutionary trends at $7\times10^3\, \mathrm{AU}$, $10^4\, \mathrm{AU}$, $5\times10^4\, \mathrm{AU}$, and $10^5\, \mathrm{AU}$ in both models resemble those plotted in Fig. \ref{figure3}, but with deviations by a factor of $3-5$. Nonetheless, the evolutions at $10^4\, \mathrm{AU}$ in model 4 and at $10^5\, \mathrm{AU}$ in model 6 can still roughly fit the observations at similar aforementioned timescales in each stage. In contrast, the evolutions at $10^2\, \mathrm{AU}$ and $10^3\, \mathrm{AU}$ exhibit greater fluctuations, which can deviate from the observations by several orders of magnitude and can be neglected in our discussion.

        \begin{table*}
        \caption{Values of $\rho$, $A_{\mathrm{V}}$, and $T$ during each stage at two different radii (10$^4$ AU and 10$^5$ AU) from the center in the best-fit models (model 4 and model 6). }
        \label{table4}
        \begin{center}
          \begin{tabular}{lcccc}\\
          \hline
          \hline
           \multirow{2}{*}{\makecell{Physical parameter}} &
            \multicolumn{2}{c}{Model 4} &
            \multicolumn{2}{c}{Model 6}  \\
            \cline{2-3}
            \cline{4-5}
            &$10^4\,\mathrm{AU}$&$10^5\,\mathrm{AU}$&$10^4\,\mathrm{AU}$&$10^5\,\mathrm{AU}$\\
            \hline
            HMSC stage&&&&\\
            $\rho$ (cm$^{-3}$)&$1.40\times10^5$&$6.34\times10^3$&$3.13\times10^5$&$4.95\times10^3$\\
            $A_{\mathrm{V}}$ (mag)&46.1&5.85&78.7&4.97\\
            $T$ (K)&11.3&11.3&20.9&20.9\\
            \hline
            HMPO stage&&&&\\
            $\rho$ (cm$^{-3}$)&$3.97\times10^5$&$6.29\times10^3$&$3.97\times10^5$&$6.29\times10^3$\\
            $A_{\mathrm{V}}$ (mag)&92.5&5.83&92.5&5.83\\
            $T$ (K)&12.2&4.84&36.5&14.5\\
            \hline
            HMC stage&&&&\\
            $\rho$ (cm$^{-3}$)&$1.69\times10^6$&$1.69\times10^4$&$1.76\times10^6$&$4.42\times10^4$\\
            $A_{\mathrm{V}}$ (mag)&228&11.2&249&21.4\\
            $T$ (K)&68.4&27.2&55.5&25.3\\
            \hline
            UCH{\sc ii} stage&&&&\\
            $\rho$ (cm$^{-3}$)&$1.06\times10^6$&$1.06\times10^4$&$6.52\times10^6$&$1.64\times10^5$\\
            $A_{\mathrm{V}}$ (mag)&178&8.26&596&51.2\\
            $T$ (K)&39.2&15.6&61.7&28.2\\
            \hline\\
          \end{tabular}\\
        \end{center}
      \end{table*}

   \par{}Subsequently, we mainly present the densities, the visual extinctions, and the temperatures at large radii of $10^4\, \mathrm{AU}$ and $10^5\, \mathrm{AU}$ in models 4 and 6, which are summarized in Table \ref{table4}. Firstly, these parameters cannot alter the overall evolutionary trends of HC$_{3}$N/N$_{2}$H$^{+}$, HC$_{3}$N/H$_{2}$, and N$_{2}$H$^{+}$/H$_2$ in these two models, especially the results at the same radius (see Fig. \ref{figure4}). Next, according to the chemical kinetics, the density primarily influences the overall evolutionary efficiency in the gas phase, while the visual extinction mainly affects the synthesis of ions via photoionization and photodissociation, especially at low values (e.g., $<5$ mag). Correspondingly, for the evolution during the HMSC stage with the same temperatures in the same model, a higher density at $10^4\, \mathrm{AU}$ induces a more rapid evolution of HC$_{3}$N/N$_{2}$H$^{+}$ and HC$_{3}$N/H$_{2}$, with larger fluctuations compared with those at $10^5\, \mathrm{AU}$, whereas N$_{2}$H$^{+}$/H$_2$ at both locations evolves steadily. In addition, a higher temperature at $10^4\, \mathrm{AU}$ or $10^5\, \mathrm{AU}$ in model 6 induces lower HC$_{3}$N/N$_{2}$H$^{+}$ and HC$_{3}$N/H$_{2}$, but higher N$_{2}$H$^{+}$/H$_2$ during the HMSC stage compared with model 4 (20.9 K vs 11.3 K), accompanied by compatible densities and visual extinctions. The discrepancies of these ratios at the same locations in the two models become more distinct during the HMPO stage, as the temperatures in model 6 are higher than those in model 4 by a factor of 3 (36.5 K vs 12.2 K at $10^4\, \mathrm{AU}$ and 14.5 K vs 4.8 K at $10^5\, \mathrm{AU}$). The abrupt increase in HC$_{3}$N/N$_{2}$H$^{+}$ and HC$_{3}$N/H$_{2}$ at the start of the HMPO and HMC stage corresponds to rapidly produced HC$_3$N, which could be induced by the sublimation of CH$_4$ from grains at $T\sim25\,\mathrm{K}$ followed by gaseous reactions (see Sect. \ref{section4.1}). This may indicate that the temperature is more significant than the density and the visual extinction to the evolution of HC$_{3}$N/N$_{2}$H$^{+}$ in HMSFRs. For the two best-fit cases, the densities and the visual extinctions at $10^4\, \mathrm{AU}$ in model 4 can be over one order of magnitude higher than those at $10^5\, \mathrm{AU}$ in model 6, but the temperatures are similar except for the HMC stage. The density during each stage probably mainly affects the evolutionary timescale for replicating the observations. The visual extinction mainly influences the evolution of ions via photodissociation in the outer region (several $10^4\,\mathrm{AU}-10^5\,\mathrm{AU}$) with low $A_{\mathrm{V}}\sim5$ during the HMSC stage and the HMPO stage. Consequently, a relatively cold diffuse envelope throughout the entire evolution of HMSFRs is essential to replicate the observed ratio of HC$_3$N/N$_2$H$^+$, while the inner regions characterized by much higher temperatures are not necessary. Such an analysis suggests that the cold envelope could play a crucial role in maintaining specific chemical conditions.

  \section{Discussion} \label{section4}
  
        \subsection{Key mechanisms affecting HC$_3$N/N$_2$H$^+$} \label{section4.1}
  
   \par{}According to derived physical conditions at the outermost radius in model  6, we discuss the key mechanisms for influencing HC$_{3}$N/N$_{2}$H$^{+}$ in HMSFRs. The major reactions for producing HC$_3$N are:
     \begin{align}
         &\mathrm{HC_3NH^+ + e^- \rightarrow HC_3N + H, } \label{reaction1} \\
         &\mathrm{CN + C_2H_2 \rightarrow HC_3N + H, } \label{reaction2} \\
         &\mathrm{N + C_3H_3 \rightarrow HC_3N + H_2. } \label{reaction3} 
     \end{align}
Meanwhile, the important reactions for destroying HC$_3$N are:
     \begin{align}
         &\mathrm{C^+ + HC_3N \rightarrow C_3H^+ + CN, } \label{reaction4} \\
         &\mathrm{C^+ + HC_3N \rightarrow C_4N^+ + H, } \label{reaction5} \\
         &\mathrm{H_3^+ + HC_3N \rightarrow HC_3NH^+ + H_2, } \label{reaction6} \\
         &\mathrm{CH_5^+ + HC_3N \rightarrow HC_3NH^+ + CH_4. } \label{reaction7}
     \end{align}
Therefore, the evolution of HC$_{3}$N in HMSFRs can be explained as follows. In the relatively cold HMSC stage with $T<25\,\mathrm{K}$, in the outer diffuse region (several $10^4\,\mathrm{AU}-10^5\,\mathrm{AU}$), the ion-molecule reactions \ref{reaction4}, \ref{reaction5}, and \ref{reaction6} mainly destroy HC$_3$N, while the dissociative recombination reaction \ref{reaction1} mainly converts the ion HC$_3$NH$^+$ back to HC$_3$N. These reactions can be further enhanced by more ions (especially C$^+$) and electrons induced by photodissociation due to the low $A_{\mathrm{V}}$. For the inner dense region ($r<10^4\,\mathrm{AU}$), the neutral-neutral reactions \ref{reaction2} and \ref{reaction3} are more significant for the production of HC$_{3}$N. They are triggered by carbon-chain species, C$_2$H$_2$ and C$_3$H$_3$, produced via the reaction of C$^+$ with H$_2$ and subsequent reactions \citep{2013ChRv..113.8981S}. In addition, reactions \ref{reaction6} and \ref{reaction7} play a major role in the destruction of HC$_{3}$N in the dense region, but most products can be converted back to HC$_{3}$N via reaction \ref{reaction1}. When $T>25\,\mathrm{K}$ during the HMPO, HMC, and UCH{\sc ii} stage, reactions \ref{reaction2} and \ref{reaction3} still remain significant, but driven by the thermal desorption of CH$_4$ and subsequent reactions with C$^+$ for synthesizing C$_2$H$_2$ and C$_3$H$_3$ \citep{2008ApJ...674..984A, 2008ApJ...681.1385H, 2011ApJ...743..182H, 2013ChRv..113.8981S, 2019ApJ...881...57T}, also known as warm carbon-chain chemistry \citep[WCCC;][]{2008ApJ...672..371S, 2009ApJ...697..769S, 2013ChRv..113.8981S, 2019A&A...622A.185W}. For even hotter regions with $T>90\,\mathrm{K}$, the sublimation of HC$_{3}$N molecules trapped on grains is the only dominating mechanism for increasing gaseous HC$_{3}$N molecules. Thus, at the start of the HMPO and HMC stage, the average abundance of HC$_3$N increases, as CH$_4$ and HC$_3$N molecules evaporate from the grains in specific regions where $T$ becomes higher than their sublimation temperatures. During the HMPO stage, the sublimation of CH$_4$ and the WCCC mechanism mainly occur within the range of $10^3-3\times10^4\,\mathrm{AU}$, which is smaller than that during the HMC stage within $3\times10^3-10^5\,\mathrm{AU}$. In contrast, the sublimation of HC$_3$N occurs in similar regions during both stages ($<10^3\,\mathrm{AU}$ during the HMPO stage and $<3\times10^3\,\mathrm{AU}$ during the HMC stage, respectively). A larger warm region during the HMC stage results in a higher increase in HC$_3$N compared to that during the HMPO stage. However, reactions \ref{reaction4} $-$ \ref{reaction7} gradually decrease the abundance of HC$_3$N over time during each stage, since CH$_4$ and HC$_3$N molecules trapped on the grains are finite.

   \par{}For N$_2$H$^+$, there is only one major formation reaction, 
     \begin{align}
         &\mathrm{H_3^+ + N_2 \rightarrow N_2H^+ + H_2, } \label{reaction8} 
     \end{align}
and the major destruction reactions are: 
     \begin{align}
         &\mathrm{N_2H^+ + e^- \rightarrow NH + N, } \label{reaction9} \\
         &\mathrm{N_2H^+ + e^- \rightarrow N_2 + H, } \label{reaction10} \\
         &\mathrm{N_2H^+ + CO \rightarrow HCO^+ + N_2, } \label{reaction11} \\
         &\mathrm{N_2H^+ + CH_4 \rightarrow CH_5^+ + N_2, } \label{reaction12} \\
         &\mathrm{N_2H^+ + NH_3 \rightarrow NH_4^+ + N_2, } \label{reaction13} \\
         &\mathrm{N_2H^+ + H_2O \rightarrow H_3O^+ + N_2. } \label{reaction14} 
     \end{align}
During the HMSC stage, $T$ is already higher than the sublimation temperature of N$_2$ (about 19 K), and sublimated N$_2$ molecules drive the formation of N$_2$H$^+$ via reaction \ref{reaction8}. In contrast, reactions \ref{reaction9} and \ref{reaction10} mainly destroy N$_2$H$^+$, driven by intense cosmic-ray ionization and photodissociation. During the hotter HMPO, HMC, and UCH{\sc ii} stages, reaction \ref{reaction8} remains significant for forming N$_2$H$^+$. But reactions \ref{reaction11}, \ref{reaction12}, \ref{reaction13}, and \ref{reaction14} can sequentially become important for the destruction of N$_2$H$^+$ when the temperature rises above $\sim21\,\mathrm{K}$, the sublimation temperature of CO, $\sim25\,\mathrm{K}$ for CH$_4$, $\sim115\,\mathrm{K}$ for NH$_3$, and $\sim120\,\mathrm{K}$ for H$_2$O, respectively. Consequently, at the start of the HMPO, HMC, and UCH{\sc ii} stages, the thermal desorption of CO, CH$_4$, NH$_3$, and H$_2$O from the grains enhances reactions \ref{reaction11} $-$ \ref{reaction14} and induces the decrease in N$_2$H$^+$ compared to that at the end of the previous stage. However, the efficiency of reaction \ref{reaction8} always overwhelms the total efficiency of reactions \ref{reaction9} $-$ \ref{reaction14} during each stage, which accumulates more N$_2$H$^+$ ions for increasing the abundance during each stage. As a result, combined with the previous analysis of HC$_3$N, the ratio HC$_3$N/N$_2$H$^+$ can rapidly increase at the start of each stage, accompanied by a decreasing trend as the stages progress. In summary, the evolution of HC$_3$N is mainly affected by the WCCC mechanism and its thermal desorption, whereas the evolution of N$_2$H$^+$ is primarily influenced by the thermal desorption of N$_2$, CO, CH$_4$, NH$_3$, and H$_2$O from the grains, followed by dissociative recombination and ion-molecule reactions.

        \subsection{Other candidate chemical clocks} \label{section4.2}
  
  \par{}Based on the evolution of HC$_3$N and N$_2$H$^+$ during different stages of HMSFRs, carbon-chain species are affected by the WCCC mechanism and their thermal desorption just like HC$_3$N, while ions are mainly influenced by cosmic-ray ionization, photodissociation, and dissociative recombination reactions as N$_2$H$^+$. Thus, we suggest that the ratio between one carbon-chain species and an ion could be a candidate chemical clock. Other candidate chemical clocks derived from observations (see Sect. \ref{section1}) can also be further analyzed with our simulations. Therefore, we examined in total 350 ratios involving 27 species, including carbon-chain species (C$_2$H, C$_3$H, C$_4$H, C$_5$H, C$_2$H$_2$, C$_3$H$_2$, C$_4$H$_2$, C$_5$H$_2$, CH$_3$CCH, HC$_3$N, HC$_5$N, and C$_2$S), simple molecules (CS, NH$_3$, HCN, HNC, CO, SO, H$_2$CO, and HNCO), COMs (HCOOH, CH$_3$OH, and CH$_3$CN), and ions (N$_2$H$^+$, HCO$^+$, H$_3^+$, and C$^+$). Except for the aforementioned HC$_3$N/N$_2$H$^+$, 178 ratios exhibit increasing or decreasing evolutionary trends at the late HMSC stage, the early HMPO stage, and the early UCH{\sc ii} stage according to our simulations. Figure \ref{appendixB_figure1} depicts the evolution of these average ratios within $10^5\, \mathrm{AU}$ from the center adopting model 6, which could be proposed as candidates of chemical clocks from the perspective of chemical evolution. Among them, 157 ratios are observable and can be verified with observations, whereas H$_3^+$ is not observable, and its related ratios cannot be considered as chemical clocks. Thus, such candidate ratios are presented as a foundational database for future observational analysis and simulations. Besides the aforementioned mechanisms affecting carbon-chain species and ions, respectively, for simple molecules, thermal desorption and ion-molecule reactions are significant, and for COMs, two-body reactions on the grains and thermal desorption are important. Nevertheless, the specific reactions affecting the evolution of each ratio can be further discussed in a future study. 
  
  \par{}According to our results, candidate chemical clocks C$_2$H/HCO$^+$ and HC$_3$N/HNC show similar evolutionary trends as observations \citep{2015MNRAS.451.2507Y, 2019MNRAS.484.4444U}, which may be further confirmed as chemical clocks. In addition, CS/SO, HCN/HNCO, C$_2$H/HNCO, and SO/N$_2$H$^+$ exhibit similar evolutionary trends compared with previous simulations \citep{2014A&A...563A..97G} as candidate chemical clocks. In contrast, C$_2$H/N$_2$H$^+$ and C$_2$H/C$_3$H$_2$ display a decreasing tendency along with the evolution, which contradicts the statistics of observations \citep{2016PASA...33...30R, 2019MNRAS.484.4444U}, although the former one, C$_2$H/N$_2$H$^+$, is consistent with previous simulations \citep{2014A&A...563A..97G}. Except for these candidate chemical clocks based on our simulations, we also depict some ratios in Fig. \ref{appendixC_figure1} that have been suggested as chemical clocks in previous studies. The ratio of N$_2$H$^+$/HCO$^+$ is indeed lower during the UCH{\sc ii} stage than that during previous stages and in agreement with observations \citep{2015MNRAS.451.2507Y, 2019MNRAS.484.4444U}, but the ratio during the HMSC stage and HMPO stage does not show clear differences. For C$_3$H$_2$/C$_2$S, NH$_3$/C$_2$S, and N$_2$H$^+$/CS, they increase from the HMSC stage to the HMPO stage and are consistent with observations \citep{2014PASJ...66..119O, 2014PASJ...66...16T}, while NH$_3$/HC$_3$N decreases from the HMSC stage to the HMPO stage and contradicts the observations \citep{2014PASJ...66..119O}. However, these four ratios apparently do not exhibit monotonic evolutionary trends when one examines their values across the HMSC, HMPO, and UCH{\sc ii} stages. For HCN/HNC, no obvious difference appears during the whole evolution, which is not consistent with the increased ratio that is observed \citep{2015ApJS..219....2J, 2016PASA...33...30R, 2019MNRAS.484.4444U} or with previous simulations \citep{2014A&A...563A..97G}. Also, the evolutionary trends of HCO$^+$/HNC, HNC/N$_2$H$^+$, HCN/N$_2$H$^+$, and CS/N$_2$H$^+$ contradict observations \citep{2016PASA...33...30R, 2019MNRAS.484.4444U}. Such deviations between simulations and observations may be driven by the simplification of the physical models adopted in the simulations. We also compared CH$_3$CN/C$_2$H, CH$_3$OH/CO, CH$_3$OH/C$_2$H, CS/C$_2$H, H$_2$CO/CO, N$_2$H$^+$/CO, and CH$_3$OH/CH$_3$CN with simulations presented by \citet{2014A&A...563A..97G}. Among them, CH$_3$CN/C$_2$H, CS/C$_2$H, N$_2$H$^+$/CO, and H$_2$CO/CO present roughly similar evolutionary trends across four stages, along with deviations in their values during the same stage. CH$_3$OH/CO, CH$_3$OH/C$_2$H, and CH$_3$OH/CH$_3$CN exhibit opposite evolutionary trends, accompanied by severe deviations in their values. Since most ratios do not exhibit monotonic evolutionary trends, it is not necessary to systematically analyze the reasons inducing the discrepancies between our simulations and results derived by \citet{2014A&A...563A..97G}. We propose that the discrepancies could be attributed to the different physical parameters between our best-fit model and previous models. Such a detailed analysis could be conducted in a future study. 
  
  \par{}Given the deviations between observations and simulations, some improvements could be considered in the future analysis for our understanding of HMSFRs. Firstly, the 1D physical model of HMSFRs, with constant parameters during each stage and abrupt transitions between two stages, is apparently not sufficiently accurate, although the chemistry can quickly respond to the discontinuities of physical parameters \citep{2014A&A...563A..97G}. A more sophisticated physical model that incorporates varying parameters and smoother transitions between stages should be developed. An example model is described by \citet{2021A&A...652A..71S}, which includes a collapse stage and a warm-up stage with continuity between them. The primary distinction between our optimal model 6 and models proposed by \citet{2014A&A...563A..97G} or \citet{2021A&A...652A..71S}  is that our model 6 adopts higher densities and temperatures within the range of $10^5\, \mathrm{AU}$, along with an enhanced cosmic ray ionization rate. Such a physical model could be further analyzed by adopting a larger range of parameters and supplementing the UCH{\sc ii} stage in a future study. Secondly, the chemical model also needs to be further modified by including more updated reactions and mechanisms, especially those related to the grain surfaces. Finally, when comparing simulations with observations, it is important to recognize that only one ratio between two species may not rigorously constrain the evolutionary stage of HMSFRs. This conclusion is mainly based on the results in panel C of Fig. \ref{figure2} (also see panel C of Fig. \ref{appendixA_figure3} and panel C of Fig. \ref{appendixA_figure4}). The average ratio of HC$_3$N/N$_2$H$^+$ within $5\times10^4\, \mathrm{AU}$ from the center can also replicate similar trends and values in the HMSC, HMPO, and UCH{\sc ii} stages, respectively. However, the average abundance ratio of HC$_3$N/H$_2$ obviously deviates from the observations. If only one ratio between two species were compared, such a discrepancy could go unnoticed. Also, such a ratio could be replicated in different stages in simulations, accompanied by different magnitudes of the abundances of the involved species. Therefore, to rigorously constrain the evolutionary stage of HMSFRs, not only the column density ratio between two species but also the abundances of these two species should be studied. Alternatively, more than two column density ratios should be simultaneously analyzed when the column density of H$_2$ is not derived.

  \section{Conclusions} \label{section5}
  
  \par{}In this paper, we have performed chemical simulations to reproduce the observed column density ratio of HC$_3$N/N$_2$H$^+$ and the abundances of these two species across various evolutionary stages in HMSFRs. Observations show an increasing evolutionary trend of HC$_3$N/N$_2$H$^+$, which is considered a chemical clock. Simultaneously, we have identified the chemical processes responsible for the observed time-dependent trends in these stages. Our simulations adopted the astrochemical code Nautilus and the existing 1D models of HMSFRs proposed by \citet{2014A&A...563A..97G, 2015A&A...579A..80G}, which include four evolutionary stages with constant densities and temperatures during each stage, as well as abrupt transitions between adjacent stages. However, both density and temperature fluctuate throughout the entire evolution. To maintain a steady increase in density and temperature over time as the global hierarchical collapse scenario predicts \citep[e.g.,][]{2019MNRAS.490.3061V}, we adjusted parameters such as the density, temperature, and time spent in each stage. The average ratios of HC$_3$N/N$_2$H$^+$ within different ranges from the center were calculated and compared with observations, as were the average abundances of HC$_3$N and N$_2$H$^+$ with respect to H$_2$. The main conclusions are listed below:
  \begin{enumerate}
    \item When averaging over large spatial scales ($\sim10^5\,\mathrm{AU}$ from the core), the best model produced successfully matches the observed column density ratio of HC$_3$N/N$_2$H$^+$ and the abundances of the species involved at specific times for each evolutionary stage; that is, the late HMSC stage, the early HMPO stage, and the early UCH{\sc ii} stage, respectively. Assuming that thermal desorption occurs instantaneously, the time required to bring the simulated and observed ratios into agreement reflects the time necessary for chemical adjustments following jumps of physical parameters between stages. The agreement is only achieved when averaging over large spatial scales, and indicates that cluster-forming regions cannot be modeled without accounting for the extended envelope.
    \item Our simulation fails to demonstrate a continuous and consistent monotonic increase in HC$_3$N/N$_2$H$^+$ throughout the entire evolution. However, some observed ratios between adjacent stages overlap, which could be induced by observational uncertainties (such as those in deriving column densities and abundances, clump classification, and systematic effects), or indicate that the evolution of HC$_3$N/N$_2$H$^+$ may not strictly monotonically increase throughout the entire evolution; for example, because a cluster is being formed in the clump, and different parcels of gas can have different physical properties and thus a different chemistry.
    \item The evolution of HC$_3$N is mainly affected by the WCCC mechanism and its own thermal desorption, while the evolution of N$_2$H$^+$ is primarily influenced by the thermal desorption of N$_2$, CO, CH$_4$, NH$_3$, and H$_2$O on the grains, followed by gaseous dissociative recombination and ion-molecule reactions.
    \item We also examined 350 other ratios involving 27 species according to our best-fit model. We find that 178 ratios exhibit an increasing or decreasing evolutionary trend at the late HMSC stage, the early HMPO stage, and the early UCH{\sc ii} stage. Among them, 157 ratios are observable and can be compared with observations, indicating that they could be considered as candidate chemical clocks. 
\end{enumerate}    
    Overall, the results obtained from the best-fitting model timescales broadly agree with statistical estimates. This indicates that 1D models with abrupt jumps in physical parameters have reached their limits in terms of the insights they can provide. Thus, more sophisticated models and advanced approaches are necessary in future studies. For instance, physical models incorporating smoother transitions between evolutionary phases might address some of the discontinuities that cause the current models to deviate from the observational data.

  \begin{acknowledgements}
We thank our referee's constructive comments to significantly improve the quality of the manuscript. This work is supported by the National Natural Science Foundation of China (NSFC) grant No. 12303032, the Natural Science Foundation of Jiangsu Province (Grant No. BK20221163), and the Excellent Postdoctoral Talent Program of Jiangsu Province (Grant No. 2022ZB475). F.D. is supported by NSFC grant No. 12041305 and the National Key R\&D Program of China grant No. 2023YFA1608000. Y.X.W. is a member of the International Max Planck Research School (IMPRS) for Astronomy and Astrophysics at the Universities of Bonn and Cologne.
  \end{acknowledgements}

  \bibliographystyle{aa} 
  \bibliography{references}

\begin{appendix}
\onecolumn

\section{Benchmark of the chemical reaction network} \label{appendix_benchmark}

  \par
\begin{figure*}[h!]
 \centering
 \includegraphics[width=0.99\textwidth]{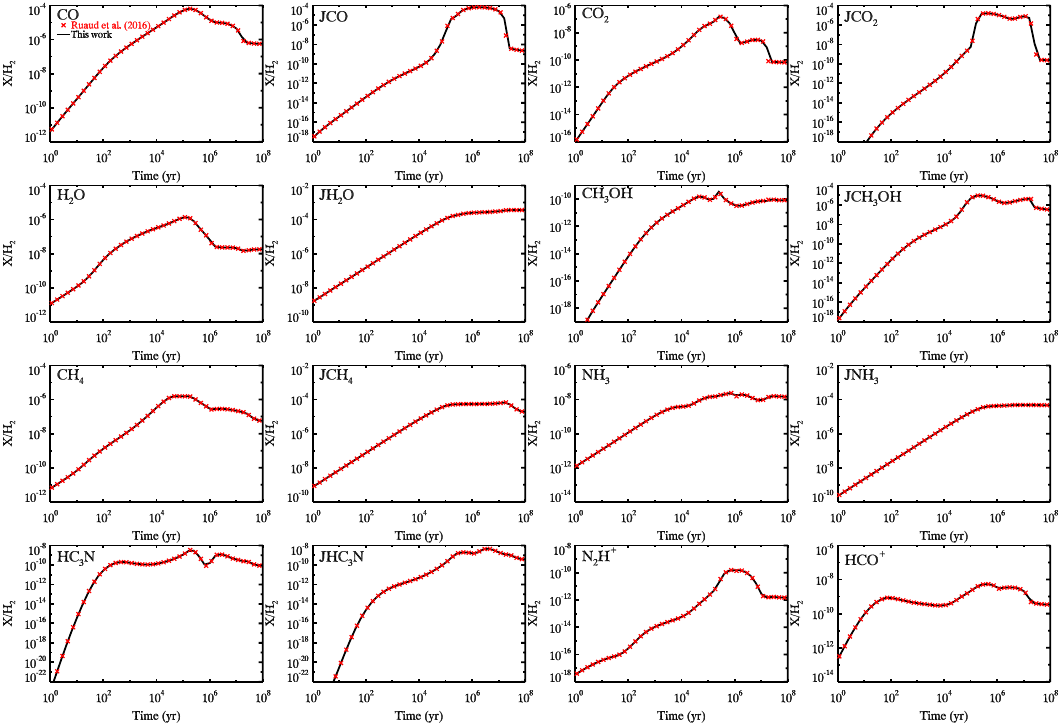}
     \caption{Time-dependent evolution of the fractional abundances of several important species (CO, CO$_2$, H$_2$O, CH$_3$OH, CH$_4$, NH$_3$, HC$_3$N, N$_2$H$^+$, and HCO$^+$) with respect to H$_2$ using the TMC1 model calculated by the Nautilus code. The letter J represents those species on the grains. The black solid lines are the results adopting the reaction network from \citet{2021A&A...648A..72W}, and the red crosses represent the results adopting the reaction network from \citet{2016MNRAS.459.3756R}, as an example network in the Nautilus package. }
     \label{appendix_figure_benchmark}
 \end{figure*}

{}Our reaction network from \citet{2021A&A...648A..72W} is benchmarked against the reaction network from \citet{2016MNRAS.459.3756R}, which is provided as an example network in the Nautilus package. The initial abundances are already listed in Table \ref{table1}. The TMC1 model proposed by \citet{2010A&A...522A..42S} is utilized in the benchmark, which is assumed as a stable physical model with the temperature $T=10\, \mathrm{K}$, the hydrogen nuclei density $n_{\mathrm{H}}=2\times10^4\, \mathrm{cm^{-3}}$, the visual extinction $A_{\mathrm{V}}=10\, \mathrm{mag}$, the cosmic ray ionization rate $\zeta_{\mathrm{CR}} = 1.8\times10^{-16} \, \mathrm{s^{-1}}$, and the FUV radiation field $\chi=1$. Other chemical parameters are the same as those described in Sect. \ref{section2}. Figure \ref{appendix_figure_benchmark} shows the time-dependent evolution of the fractional abundances of several important species with respect to H$_2$ calculated by the Nautilus code with adopting these two networks, respectively. The comparison shows good agreement between two simulations. 

\onecolumn
  \section{Additional simulated HC$_3$N/N$_2$H$^+$} \label{appendixA}

 \begin{figure*}[h!]
 \centering
 \includegraphics[width=1.0\textwidth]{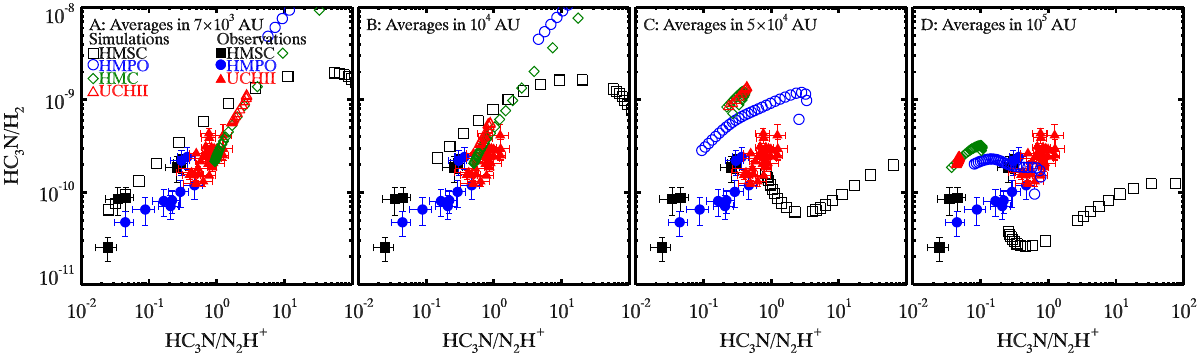}
     \caption{Same as Fig. \ref{figure1}, but adopting model 1.}
     \label{appendixA_figure1}
 \end{figure*}
 
 \begin{figure*}[h!]
 \centering
 \includegraphics[width=1.0\textwidth]{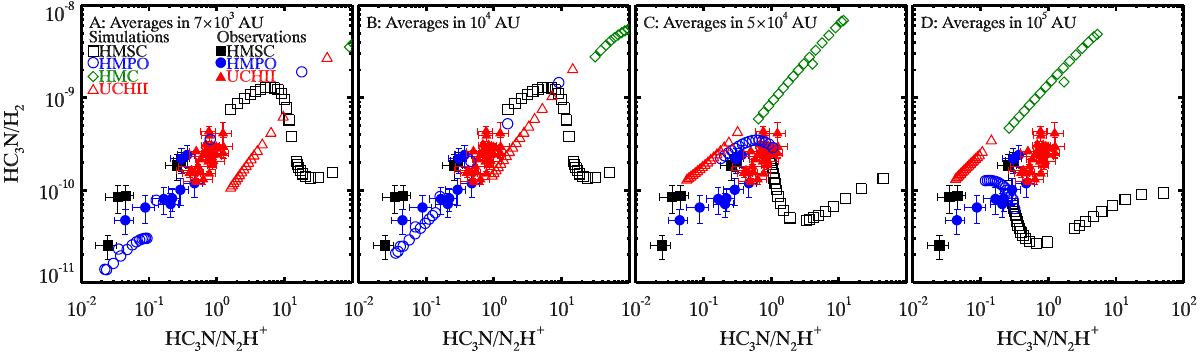}
     \caption{Same as Fig. \ref{figure1}, but adopting model 2.}
     \label{appendixA_figure2}
 \end{figure*}
 
  \begin{figure*}[h!]
 \centering
 \includegraphics[width=1.0\textwidth]{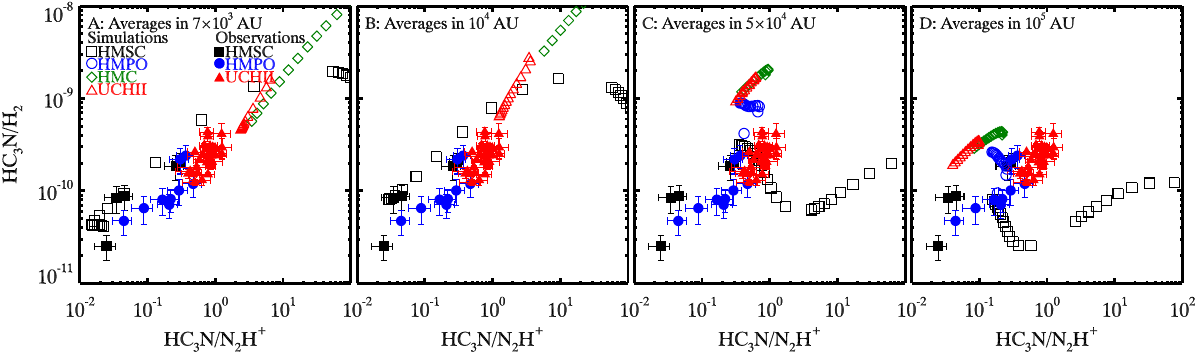}
     \caption{Same as Fig. \ref{figure1}, but adopting model 3.}
     \label{appendixA_figure3}
 \end{figure*}
 
  \begin{figure*}[h!]
 \centering
 \includegraphics[width=1.0\textwidth]{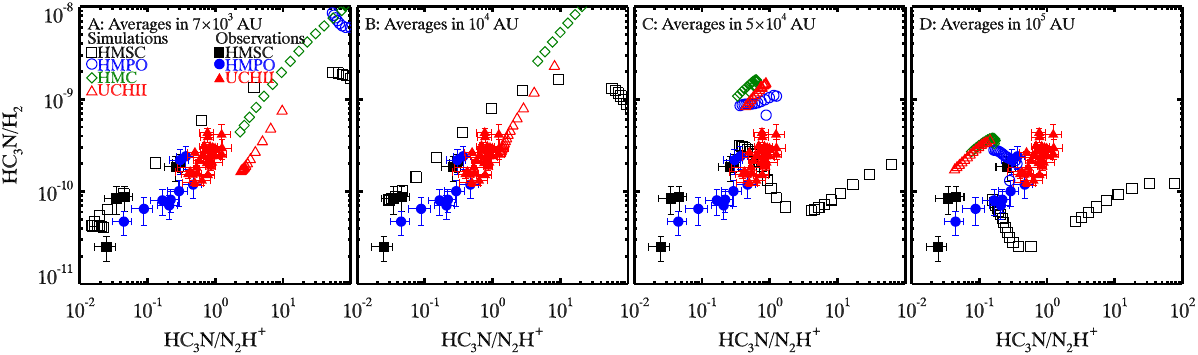}
     \caption{Same as Fig. \ref{figure1}, but adopting model 5.}
     \label{appendixA_figure4}
 \end{figure*}
 
\onecolumn 
  \section{Candidate chemical clocks} \label{appendixB} 
  
 \begin{figure*}[h!]
 \centering
  \includegraphics[width=0.71\textwidth]{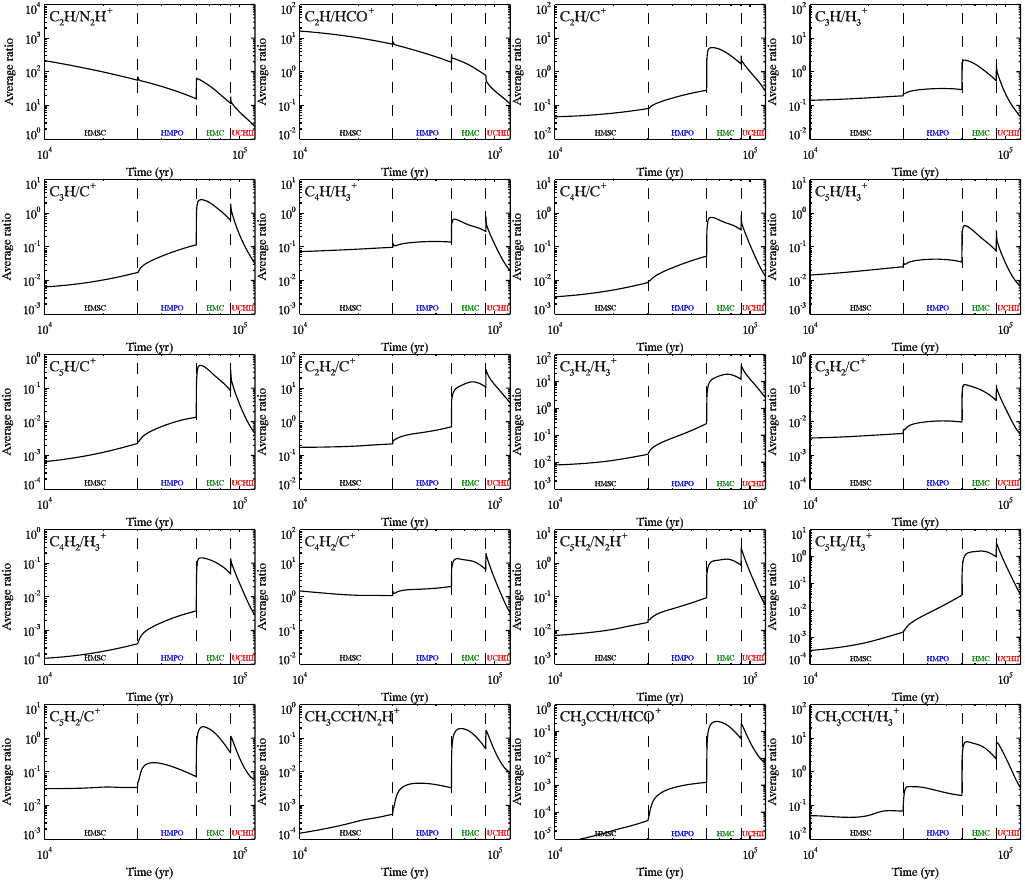}
  \includegraphics[width=0.71\textwidth]{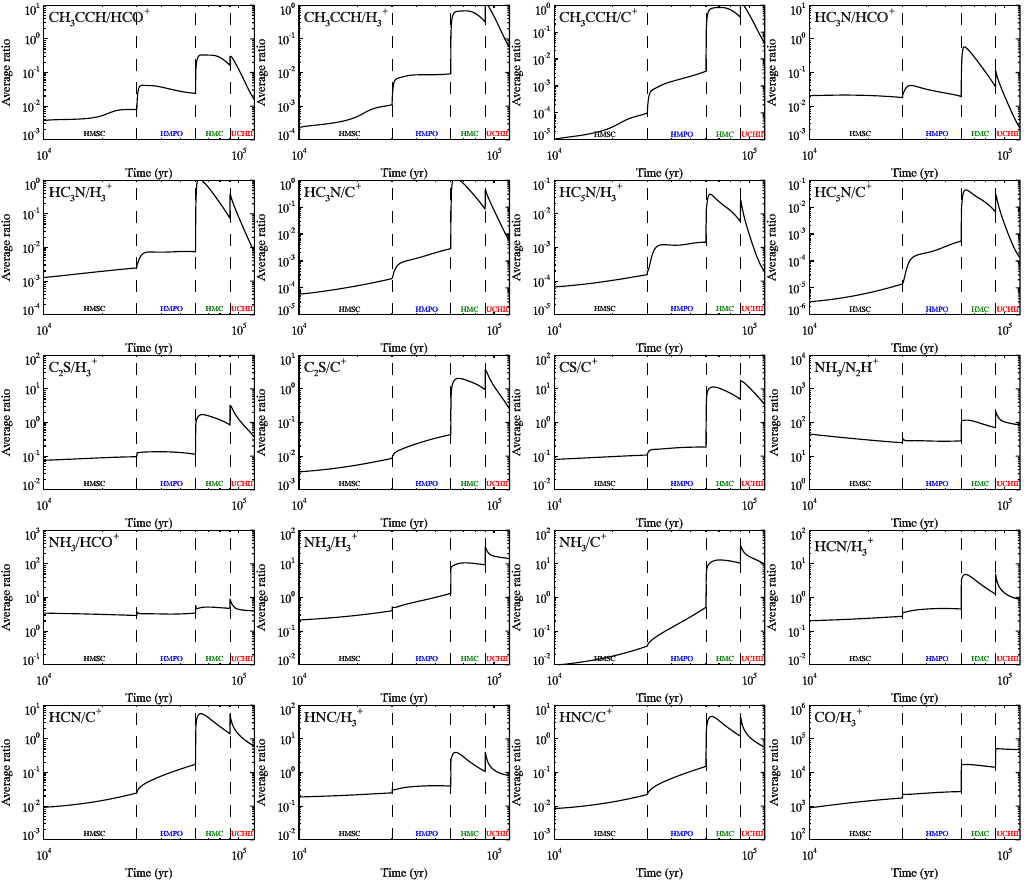}
     \caption{Time-dependent evolution of other average ratios within $10^5\, \mathrm{AU}$ from the center adopting model 6 as candidate chemical clocks. } 
     \label{appendixB_figure1}
 \end{figure*}
 
\addtocounter{figure}{-1} 
 \begin{figure*}[h!]
 \centering
 \includegraphics[width=0.71\textwidth]{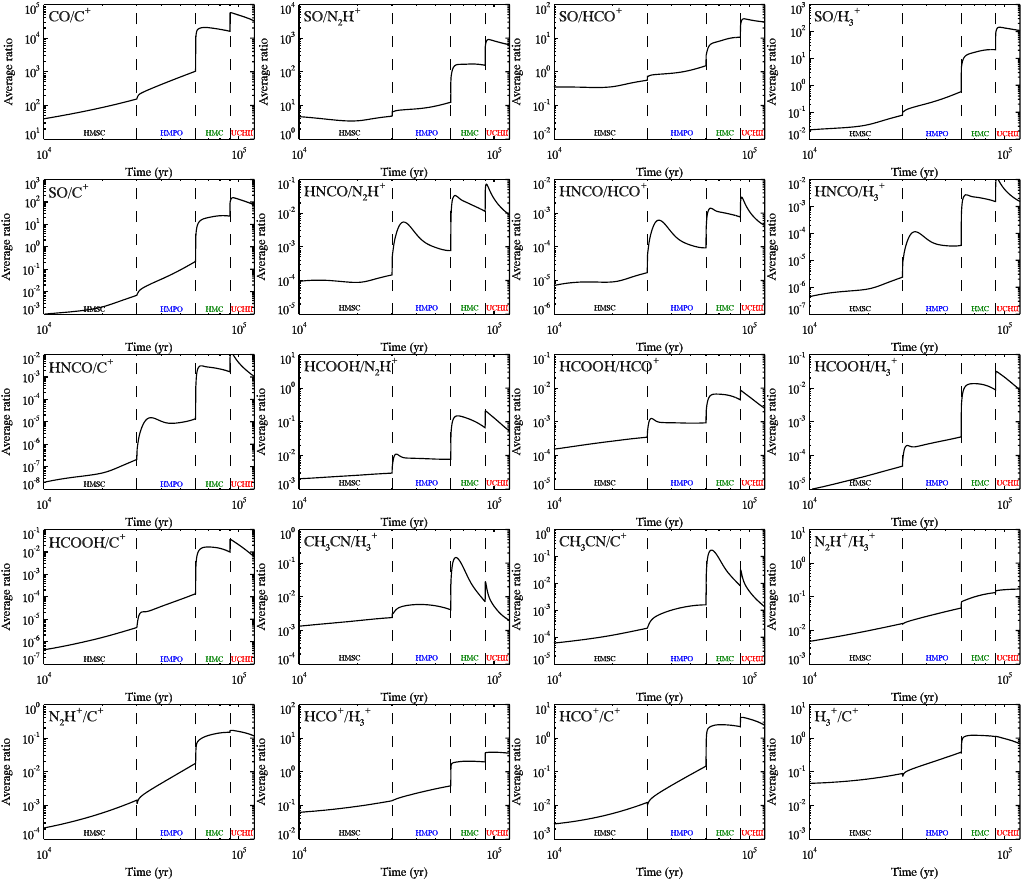}
 \includegraphics[width=0.71\textwidth]{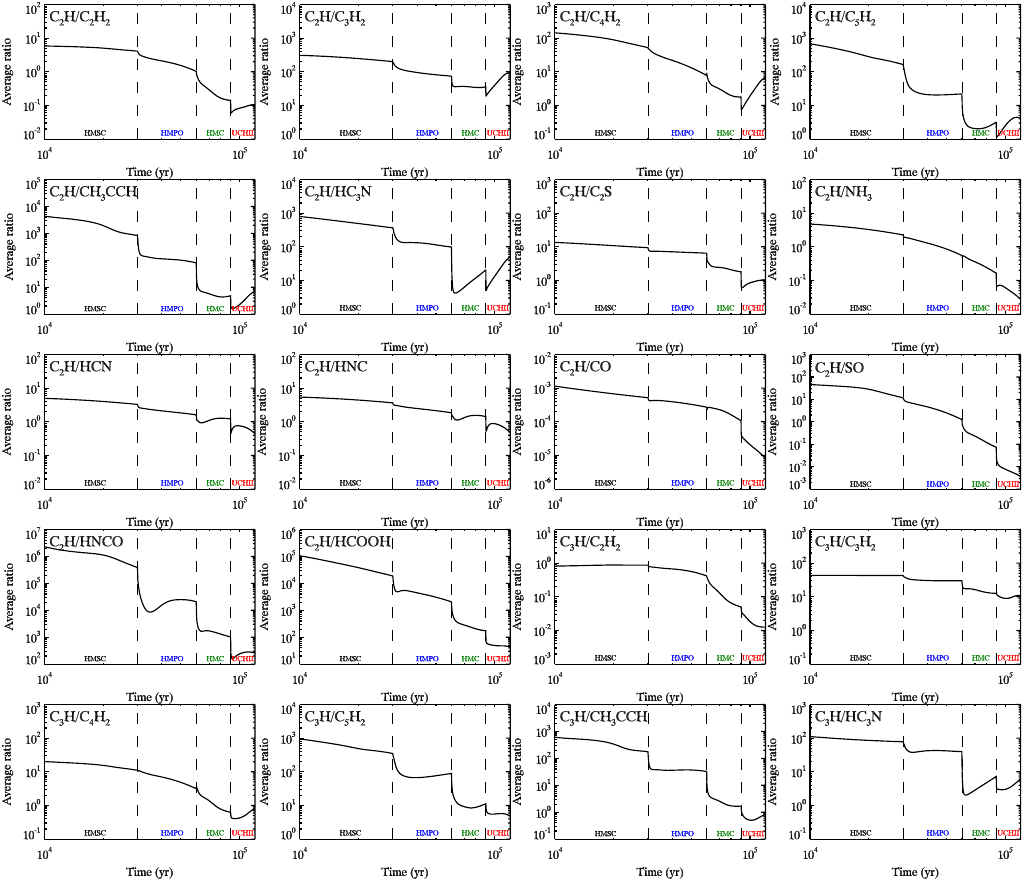}
     \caption{Continued. }
 \end{figure*}
  
 \addtocounter{figure}{-1}
 \begin{figure*}[h!]
 \centering
 \includegraphics[width=0.71\textwidth]{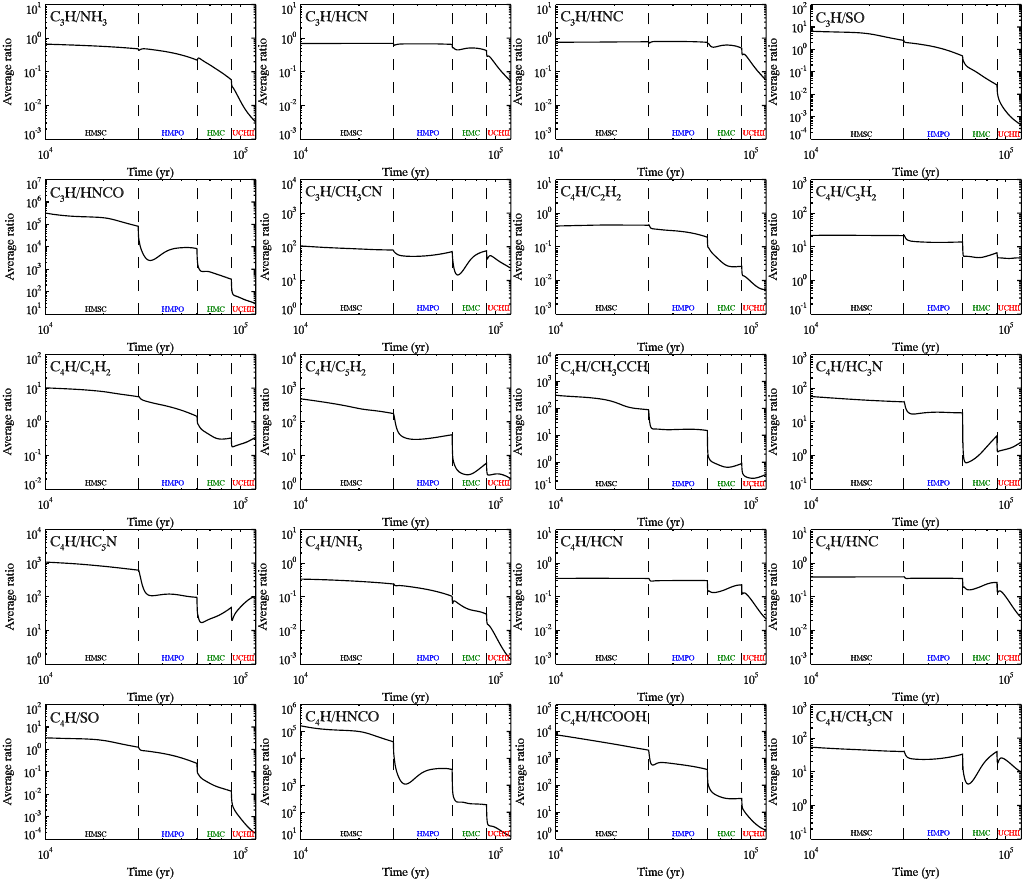}
 \includegraphics[width=0.71\textwidth]{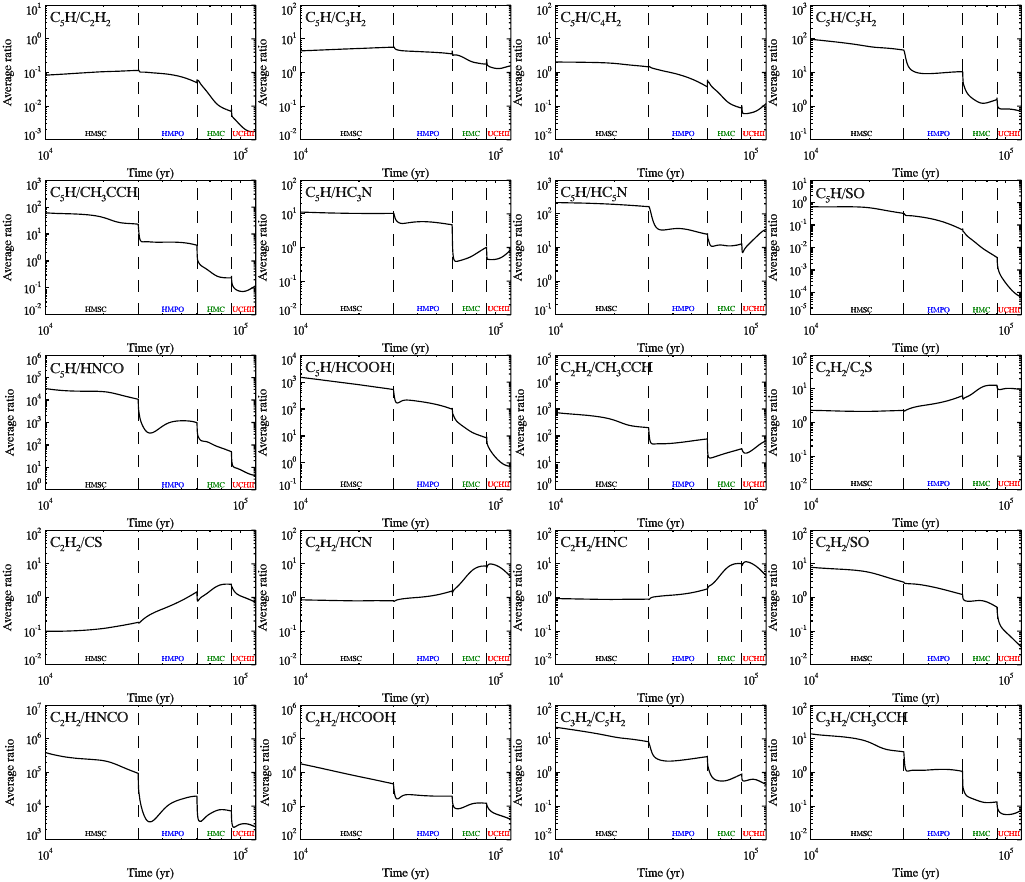}
     \caption{Continued. }
     \label{appendixB_figure2}
 \end{figure*}
 
  \addtocounter{figure}{-1}
 \begin{figure*}[h!]
 \centering
 \includegraphics[width=0.71\textwidth]{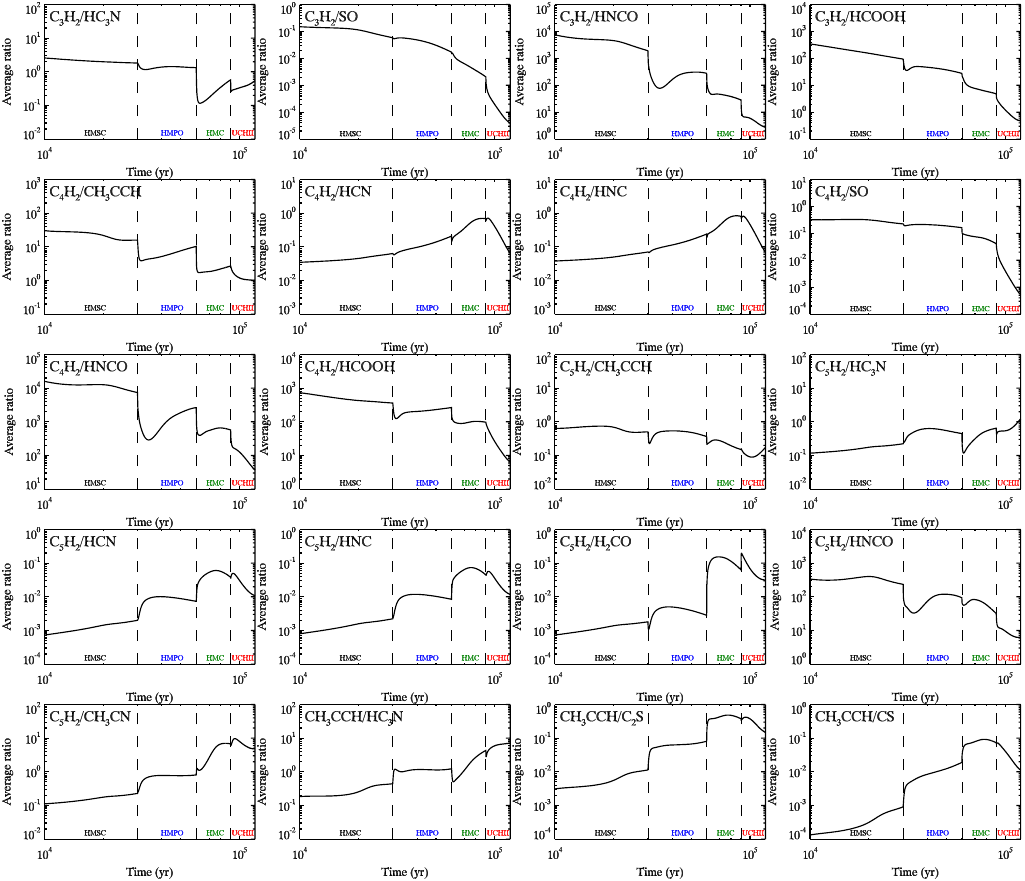}
 \includegraphics[width=0.71\textwidth]{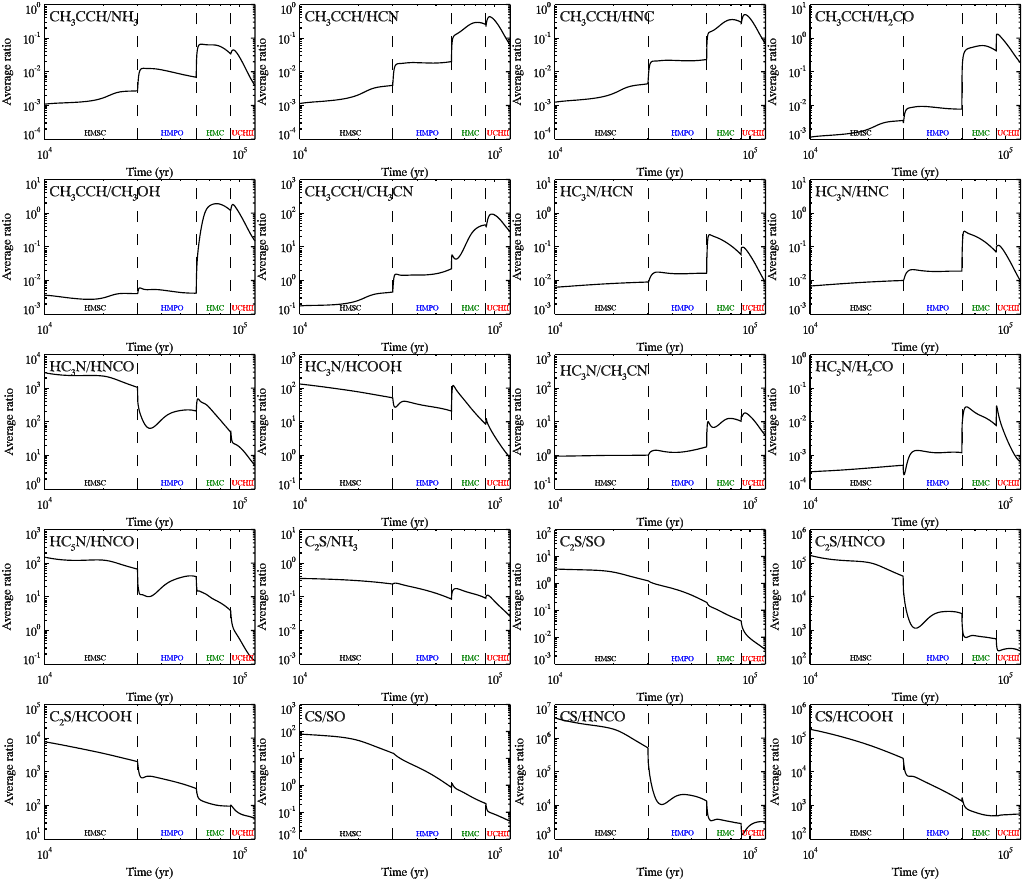}
     \caption{Continued. }
     \label{appendixB_figure3}
 \end{figure*}
 
  \addtocounter{figure}{-1}
 \begin{figure*}[h!]
 \centering
 \includegraphics[width=0.71\textwidth]{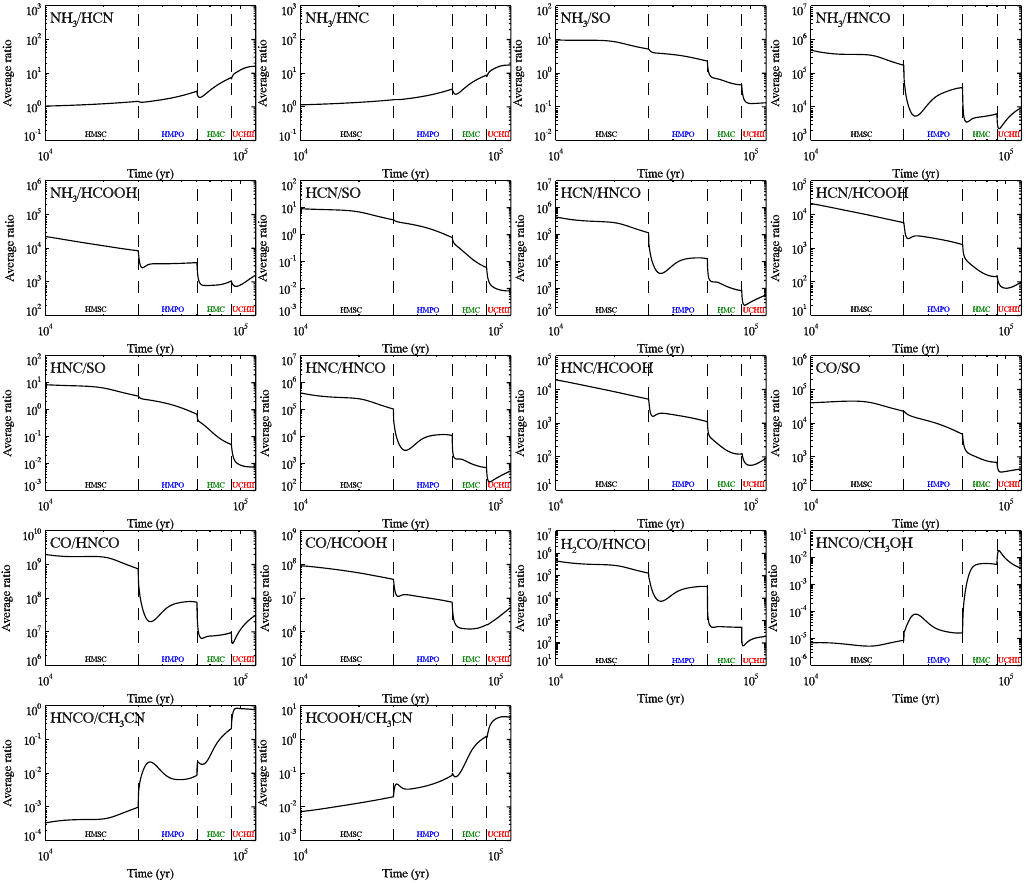}
     \caption{Continued. }
     \label{appendixB_figure4}
 \end{figure*}

 \onecolumn 
  \section{Other average ratios} \label{appendixC} 
  
\begin{figure*}[h!]
 \centering
 \includegraphics[width=0.71\textwidth]{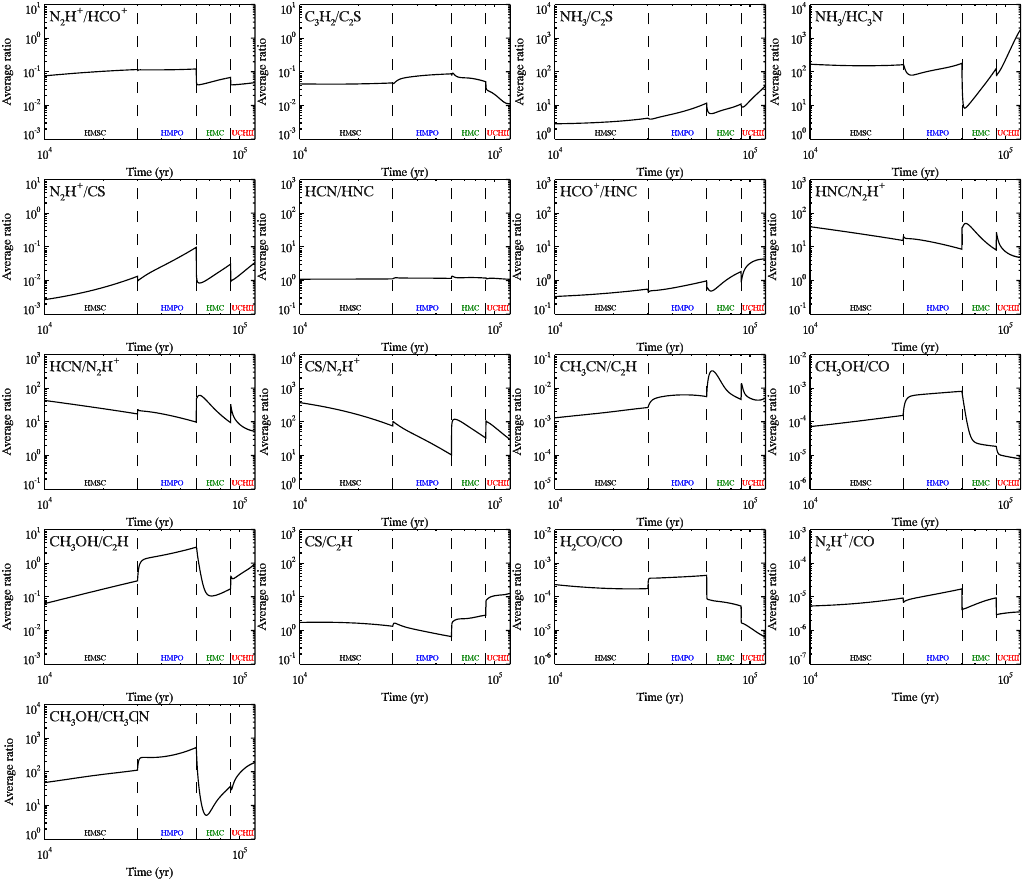}
     \caption{Other average ratios as in Fig. \ref{appendixB_figure1} but do not exhibit monotonic evolutionary trend throughout the entire evolution. }
     \label{appendixC_figure1}
 \end{figure*} 

\end{appendix}

\end{document}